\documentclass[conference]{IEEEtran}
\usepackage{cite}
\usepackage{amsmath,amssymb,amsfonts}
\usepackage{algorithmic}
\usepackage{graphicx}
\usepackage{textcomp}
\usepackage{subcaption}
\usepackage{algorithm}
\usepackage{caption}
\usepackage{array}
\newcolumntype{L}[1]{>{\raggedright\let\newline\\\arraybackslash\hspace{0pt}}m{#1}}
\newcolumntype{C}[1]{>{\centering\let\newline\\\arraybackslash\hspace{0pt}}m{#1}}
\newcolumntype{R}[1]{>{\raggedleft\let\newline\\\arraybackslash\hspace{0pt}}m{#1}}
\usepackage{array, multirow, graphicx}
\usepackage{multicol}
\usepackage{pdflscape}
\usepackage{graphicx}
\usepackage{stfloats}
\usepackage{url}
\usepackage{verbatim}
\usepackage{hyperref}
\usepackage{multirow}
\usepackage{longtable}
\usepackage{adjustbox}
\usepackage[T1]{fontenc}
\usepackage[table]{xcolor} 
\usepackage{comment}

\def\BibTeX{{\rm B\kern-.05em{\sc i\kern-.025em b}\kern-.08em
    T\kern-.1667em\lower.7ex\hbox{E}\kern-.125emX}}
\begin{document}
\title{Autonomous Vehicle Security: \\ A Deep Dive into Threat Modeling}

\author{
    \IEEEauthorblockN{Amal Yousseef\textsuperscript{1},  Shalaka Satam\textsuperscript{1}, Banafsheh Saber Latibari\textsuperscript{1},\\ Jesus Pacheco\textsuperscript{3}
     Soheil Salehi\textsuperscript{1},
    Salim Hariri\textsuperscript{1}, Partik Satam\textsuperscript{2}}
    \IEEEauthorblockA{
        \textsuperscript{1}Department of Electrical and Computer Engineering, University of Arizona, Tucson, AZ, USA\\ 
        \textsuperscript{2}Department of Systems and Industrial Engineering, University of Arizona, Tucson, AZ, USA\\
        \textsuperscript{3}Department of Industrial Engineering, University of Sonora, Hermosillo, Mexico\\
        Emails:\{\textsuperscript{1}amalyousseef, \textsuperscript{1}shalakasatam,\textsuperscript{1}banafsheh, \textsuperscript{1}ssalehi, \textsuperscript{1}hariri, \textsuperscript{2}pratiksatam\}@arizona.edu, \textsuperscript{3}jesus.pacheco@unison.mx
        }
    }

\maketitle
\begin{abstract}
    Autonomous vehicles (AVs) are poised to revolutionize modern transportation, offering enhanced safety, efficiency, and convenience. However, the increasing complexity and connectivity of AV systems introduce significant cybersecurity challenges. This paper provides a comprehensive survey of AV security with a focus on threat modeling frameworks, including STRIDE, DREAD, and MITRE ATT\&CK, to systematically identify and mitigate potential risks. The survey examines key components of AV architectures, such as sensors, communication modules, and electronic control units (ECUs), and explores common attack vectors like wireless communication exploits, sensor spoofing, and firmware vulnerabilities. Through case studies of real-world incidents, such as the Jeep Cherokee and Tesla Model S exploits, the paper highlights the critical need for robust security measures. Emerging technologies, including blockchain for secure Vehicle-to-Everything (V2X) communication, AI-driven threat detection, and secure Over-The-Air (OTA) updates, are discussed as potential solutions to mitigate evolving threats. The paper also addresses legal and ethical considerations, emphasizing data privacy, user safety, and regulatory compliance. By combining threat modeling frameworks, multi-layered security strategies, and proactive defenses, this survey offers insights and recommendations for enhancing the cybersecurity of autonomous vehicles.
\end{abstract}

\section{Introduction}\label{sec:intro}
\subsection{Background on Autonomous Vehicles}
Autonomous vehicles (AVs) represent a transformative shift in transportation technology, promising increased safety, efficiency, and convenience in modern mobility systems. These vehicles rely on a combination of sensors, machine learning, and real-time decision-making algorithms to navigate and interact with their environment without human intervention. According to the Society of Automotive Engineers (SAE), AVs are classified into six levels of automation, ranging from Level 0 (no automation) to Level 5 (full automation), with varying degrees of driver involvement and autonomy across levels \cite{sae2018taxonomy}.\\
As AVs progress toward higher levels of autonomy, their reliance on interconnected systems such as Vehicle-to-Vehicle (V2V), Vehicle-to-Infrastructure (V2I), and Vehicle-to-Everything (V2X) communication increases \cite{alotaibi2018secure}. These communication protocols allow AVs to share information, enhance situational awareness, and improve overall driving decisions. However, this interconnectivity also exposes AVs to a wide range of cybersecurity threats, making security a critical aspect of their deployment and adoption \cite{hussain2019autonomous}.
\subsection{B. Evolution of Cybersecurity Challenges in AVs}
The evolution of AV technology has been paralleled by a growing complexity in cybersecurity challenges. Early vehicles with limited connectivity relied on isolated in-vehicle networks such as the Controller Area Network (CAN), which were susceptible primarily to physical attacks \cite{miller2015remote}. As AVs adopted more sophisticated communication protocols and incorporated cloud-based and edge computing systems, they became increasingly vulnerable to remote attacks via wireless communication channels such as WiFi, Bluetooth, and cellular networks \cite{tencent2020}

Recent advancements in Artificial Intelligence (AI) and the Internet of Things (IoT) have further complicated AV security landscapes. AI-based models enhance AV capabilities in perception, decision-making, and anomaly detection but are also vulnerable to adversarial attacks that exploit system blind spots \cite{girdhar2023cybersecurity}. IoT integration has expanded attack surfaces, allowing cyber adversaries to target not only AV systems but also interconnected devices and infrastructure \cite{alotaibi2018secure}.

\subsection{Importance of Security in Autonomous Vehicles}
The security of autonomous vehicles is paramount, as vulnerabilities in AV systems can lead to catastrophic outcomes. Cyberattacks targeting AVs may compromise the vehicle's control, navigation, or decision-making systems, posing risks not only to passengers but also to other road users and infrastructure \cite{petit2015cyberattacks}. Incidents such as remote hijacking of vehicle systems, sensor spoofing, and unauthorized access to critical networks have underscored the potential dangers associated with AV security breaches \cite{ valasek2014adventures, miller2015remote}.\\
In addition to safety concerns, the financial and reputational damage to automakers and service providers can be substantial if they fail to adequately protect their AV systems from cyber threats. Furthermore, the integration of personal data and location-based services into AVs raises significant privacy concerns, which must be addressed through robust security measures \cite{hussain2019autonomous}.

\subsection{Objectives and Scope of the Survey}
This survey paper provides an in-depth review of autonomous vehicle (AV) security, with a primary focus on threat modeling to identify and categorize potential risks to AV systems. It systematically examines the threats and vulnerabilities that affect AVs, leveraging threat modeling frameworks like STRIDE and DREAD to assess the likelihood and impact of attacks on key components such as sensors, GPS systems, and communication modules. The survey also reviews security countermeasures, from cryptographic methods and secure protocols to AI-based anomaly detection systems, highlighting how each aligns with specific threat models. In addition to technical solutions, the paper discusses regulatory considerations, emphasizing the importance of proactive threat modeling in shaping secure AV standards. Covering both in-vehicle systems and external infrastructure, this survey identifies open problems and proposes future research directions focused on advancing robust, threat-model-driven security strategies for autonomous vehicles.

\subsection{Structure of the Paper}
The remainder of this paper is organized as follows:
Section \ref{sec:AVs}presents an overview of autonomous vehicle architectures, detailing key components such as sensors, communication systems, and Electronic Control Units (ECUs). Case studies of various vehicle architectures, including the Jeep Cherokee, Lexus, and Tesla, are provided to illustrate the technological underpinnings of modern AV systems.
\\
Section \ref{sec:attacks} analyzes the primary cybersecurity threats faced by AVs. It examines common attack vectors such as wireless communication exploits, sensor spoofing, and firmware vulnerabilities. Real-world case studies are included to demonstrate the practical implications of these security challenges.
\\
Section \ref{sec:modelling} introduces and evaluates prominent threat modeling frameworks, including attack-centric, asset-centric, and software-centric approaches. It also discusses methodologies such as attack trees and the MITRE ATT\&CK framework for systematically identifying and categorizing potential threats and vulnerabilities.
\\
Section  \ref{sec:strideanddread} applies the STRIDE and DREAD threat modeling frameworks to AV architectures. Threats are categorized and assessed for impact, and a comprehensive risk evaluation table is presented to prioritize mitigation strategies based on severity and exploitability.
\\
Section \ref{sec:analysis} provides a comparative analysis of STRIDE, DREAD, and MITRE ATT\&CK frameworks. The strengths, limitations, and applicability of each approach to AV security are discussed, with recommendations for integrating multiple frameworks to achieve a comprehensive threat assessment.
\\
Section \ref{sec:legal} addresses the legal and ethical challenges associated with AV security, including data privacy, user safety, transparency, and regulatory compliance. Relevant standards and regulations, such as GDPR, ISO/SAE 21434, and UNECE WP.29, are examined in the context of AV development and deployment.
\\
Section \ref{sec:future} explores emerging technologies and research directions in AV security, including blockchain-based V2X communication, AI-driven threat detection, and secure Over-The-Air (OTA) update mechanisms. Practical recommendations for AV developers, policymakers, and researchers are proposed.
\\
Section \ref{sec:conc} synthesizes the key findings and contributions of the paper, emphasizing the importance of robust cybersecurity measures and identifying future research opportunities to enhance AV security.

\section{Autonomous Vehicles}\label{sec:AVs}
This section introduces Autonomous Vehicles (AVs) and their architectures and presents a case study of three vehicles, namely, Jeep Cherokee, Lexus and Tesla,  under complex cyber-attacks.\\
Vehicles are rapidly advancing toward full autonomy by integrating sensors and vehicular networks into their control systems. The Society of Automotive Engineers (SAE), in its J3016 standard \cite{schmittner2016using, steger2016security}, defines six levels of autonomy, ranging from level 0 (no autonomy) to level 5 (complete autonomy), as outlined in Table \ref{tab:level_sae}. With the growing autonomy of vehicles, there is an increasing reliance on situational awareness, which is achieved through data-driven machine learning algorithms and artificial intelligence, powered by onboard sensor networks and external communications \cite{biswas2006vehicle, harding2014vehicle}. Policymakers predict that the global autonomous vehicle (AV) market could reach \$7 trillion by 2050, leading to enhanced transportation efficiency, fewer accidents, improved safety, and reduced environmental impact \cite{lango2022}.

\begin{table}[]
\caption{SAE Vehicle Autonomy levels}
\resizebox{\columnwidth}{!}{%
\begin{tabular}{|c|c|c|c|c|c|}
\hline
\rowcolor[HTML]{EFEFEF} 
\textbf{Level} & \textbf{Automation} & \textbf{\begin{tabular}[c]{@{}c@{}}Steering \\ Cruising\end{tabular}} & \textbf{\begin{tabular}[c]{@{}c@{}}Environment \\ Monitoring\end{tabular}} & \textbf{\begin{tabular}[c]{@{}c@{}}Fallback \\ Control\end{tabular}} & \textbf{\begin{tabular}[c]{@{}c@{}}Driving \\ Mode\end{tabular}} \\ \hline
0              & None                & H                                                                     & H                                                                          & H                                                                    & N/A                                                              \\ \hline
1              & Supportive          & H,S                                                                   & H                                                                          & H                                                                    & Some                                                             \\ \hline
2              & Partial             & S                                                                     & H                                                                          & H                                                                    & Some                                                             \\ \hline
3              & Conditional         & S                                                                     & S                                                                          & H                                                                    & Some                                                             \\ \hline
4              & High                & S                                                                     & S                                                                          & H                                                                    & Some                                                             \\ \hline
5              & Full                & S                                                                     & S                                                                          & S                                                                    & All                                                              \\ \hline
\end{tabular}%
}
\label{tab:level_sae}

\end{table}

Contrary to popular belief, most modern vehicles are equipped with connected technologies, placing them between Level 2 and Level 4 autonomy. Even basic cars today contain at least 30 Electronic Control Units (ECUs), while luxury models may have up to 100 ECUs, managing over 100 million lines of code \cite{zax2012, motavalli2010}. Each ECU \cite{kvaser_linbus} functions as a standalone computer, featuring a microprocessor and interfacing with sensors, actuators, and communication networks to perform various tasks, from operating turn signals to regulating the engine's air-fuel ratio. These ECUs communicate via different buses, such as the Controller Area Network (CAN), Ethernet, FlexRay, and Local Interconnect Network (LIN), which vary in specifications, speed, and cost depending on the control requirements.\\
In addition, modern vehicles are equipped with satellite communications (for satellite radio and navigation systems), cellular networks, WiFi, and Bluetooth. These features, typically managed through an infotainment system or head unit \cite{tesla_support}, offer drivers enhanced comfort and seamless control. Semi-autonomous vehicles also include embedded SIMs (eSIMs) that transmit telematics data to cloud servers, enabling vehicle owners to manage, track, and operate their cars using smartphone applications. These cloud-based features facilitate large-scale data collection and continual vehicle improvements, boosting performance and capabilities. The inclusion of eSIMs also allows for over-the-air (OTA) software updates, enabling manufacturers to enhance a vehicle’s functionality over time as technology evolves and components age. Researchers and automotive developers are working on Vehicle-to-Everything (V2X) communication systems, utilizing 802.11p (Wireless Local Area Network (WLAN)) and C-V2X, a cellular-based system powered by 5G and future technologies. All these technologies, integrated through the head unit, provide users with unprecedented access and control over their vehicles \cite{tencent2020, mahaffey2015,miller2015remote}
\subsection{Overview of Autonomous Vehicle Architecture}
Autonomous vehicle (AV) architectures have evolved significantly, incorporating advanced technologies such as high-performance computing, artificial intelligence, and robust communication networks. This section provides an updated overview of AV architectures, highlighting developments in sensor integration, network infrastructure, and connectivity solutions that support advanced driving functions.\\
\begin{itemize}
    \item \textbf{Multi-Sensor Fusion and High-Resolution Sensing Layer:}
Modern AVs use a complex array of sensors, including LiDAR, radar, cameras, and ultrasonic sensors, to build a comprehensive environmental model. Sensor fusion, which combines data from multiple sources, allows AVs to achieve greater accuracy and reliability in object detection and environmental awareness. Recent architectures incorporate AI-based sensor fusion algorithms that dynamically adjust to changing road conditions, reducing noise and inaccuracies in real-time \cite{ma2020artificial}.\\
\textbf{Example:} Tesla’s FSD (Full Self-Driving) architecture has evolved to process multiple video frames from an array of onboard cameras, utilizing neural networks that allow the system to predict object trajectories. By fusing radar and camera data, Tesla’s architecture minimizes reliance on LiDAR, favoring a vision-based approach to environmental perception \cite{girdhar2023cybersecurity}.

    \item \textbf{igh-Performance Computing and AI Processing Units:}
Current AVs integrate specialized high-performance computing (HPC) units designed to handle complex data processing for real-time decision-making. These HPC systems often contain GPUs, TPUs, or dedicated AI processors optimized for deep learning tasks, allowing for rapid processing of sensor data and execution of driving algorithms.\\
\textbf{Example:} Nvidia’s DRIVE AGX platform is widely used in AV architectures for its powerful AI processing capabilities, handling tasks like perception, mapping, and path planning. By incorporating real-time parallel processing, DRIVE AGX enables AVs to make instantaneous decisions, even in highly dynamic environments \cite{tabani2021performance}.

    \item \textbf{Enhanced Intra-Vehicle Network Architecture:}
The Controller Area Network (CAN) bus, traditionally used for intra-vehicle communication, is increasingly being supplemented or replaced by high-speed Ethernet networks that support the substantial data throughput required by modern AVs. Automotive Ethernet reduces latency and enhances data exchange rates, making it suitable for real-time processing and communication between critical subsystems like sensors, AI processors, and control modules.\\
\textbf{Example:} Tesla’s architecture leverages Automotive Ethernet, which allows efficient data exchange between various Electronic Control Units (ECUs) and the centralized computing unit. This setup facilitates rapid processing of high-bandwidth data, essential for advanced functionalities such as autonomous navigation and over-the-air (OTA) updates \cite{cena2023composite}.

    \item \textbf{V2X (Vehicle-to-Everything) Communication Layer:}
V2X technology is now a critical component in AV architectures, enabling AVs to interact with infrastructure (V2I), other vehicles (V2V), pedestrians (V2P), and networks (V2N). This layer enhances situational awareness and supports cooperative driving, improving traffic efficiency and safety. V2X technology is often built on 5G and Cellular V2X (C-V2X) networks, which provide the low latency and high data transfer rates required for real-time data exchange.\\
\textbf{Example:} Ford and other automakers are integrating C-V2X as part of their AV communication infrastructure, leveraging 5G for fast and secure data exchange that enables vehicles to respond proactively to road hazards and traffic signals. V2X communication also improves data sharing in complex environments, enabling better route planning and situational awareness \cite{khayyam2020artificial}.

    \item \textbf{Cloud-Based and Edge Computing Integration:1}
To manage the vast data generated by AVs, modern architectures are increasingly integrating cloud and edge computing. The cloud allows for large-scale data storage and analysis, enabling long-term learning and system updates, while edge computing provides localized data processing, reducing latency and improving responsiveness. This hybrid approach enables AVs to perform real-time decision-making with minimal delays while accessing cloud-stored data for high-level insights and updates.\\
\textbf{Example:} Waymo utilizes a cloud-edge architecture where critical, low-latency tasks are processed locally on the AV, while non-urgent data, like map updates and historical driving data, is stored and processed in the cloud. This setup allows for a balance between immediate data processing needs and centralized learning for continuous improvements \cite{naz2022intelligence}.

\item \textbf{Applications Layer: AI-Driven Decision-Making and Control:}
At the highest level, AVs use AI-driven systems to support applications like perception, localization, decision-making, and control. Advanced machine learning models process data from the sensing and network layers to understand the vehicle’s environment, predict potential hazards, and make real-time driving decisions. Reinforcement learning and neural networks are increasingly employed to optimize driving strategies, allowing AVs to learn from experience and improve over time.\\
\textbf{Example:} AI models in AVs like those used by Waymo and Uber's autonomous fleets support real-time path planning and adaptive control systems, allowing these vehicles to navigate safely and adapt to unexpected conditions on the road. These systems are designed to improve over time through self-learning mechanisms that enhance driving behavior with each trip \cite{ma2020artificial}.
\end{itemize}
The evolution of AV architectures reflects the increasing complexity of autonomous driving technology. By integrating high-speed networks, advanced AI processors, multi-sensor fusion, and robust V2X communication systems, these architectures support the computational and communication demands of modern AV systems. This layered approach enables AVs to perform intricate perception, planning, and decision-making tasks necessary for safe and efficient autonomous navigation in real-world environments.

\subsection{Example Vehicle Architectures}
\begin{figure}
\begin{center}
\includegraphics[clip,width=0.8\columnwidth]{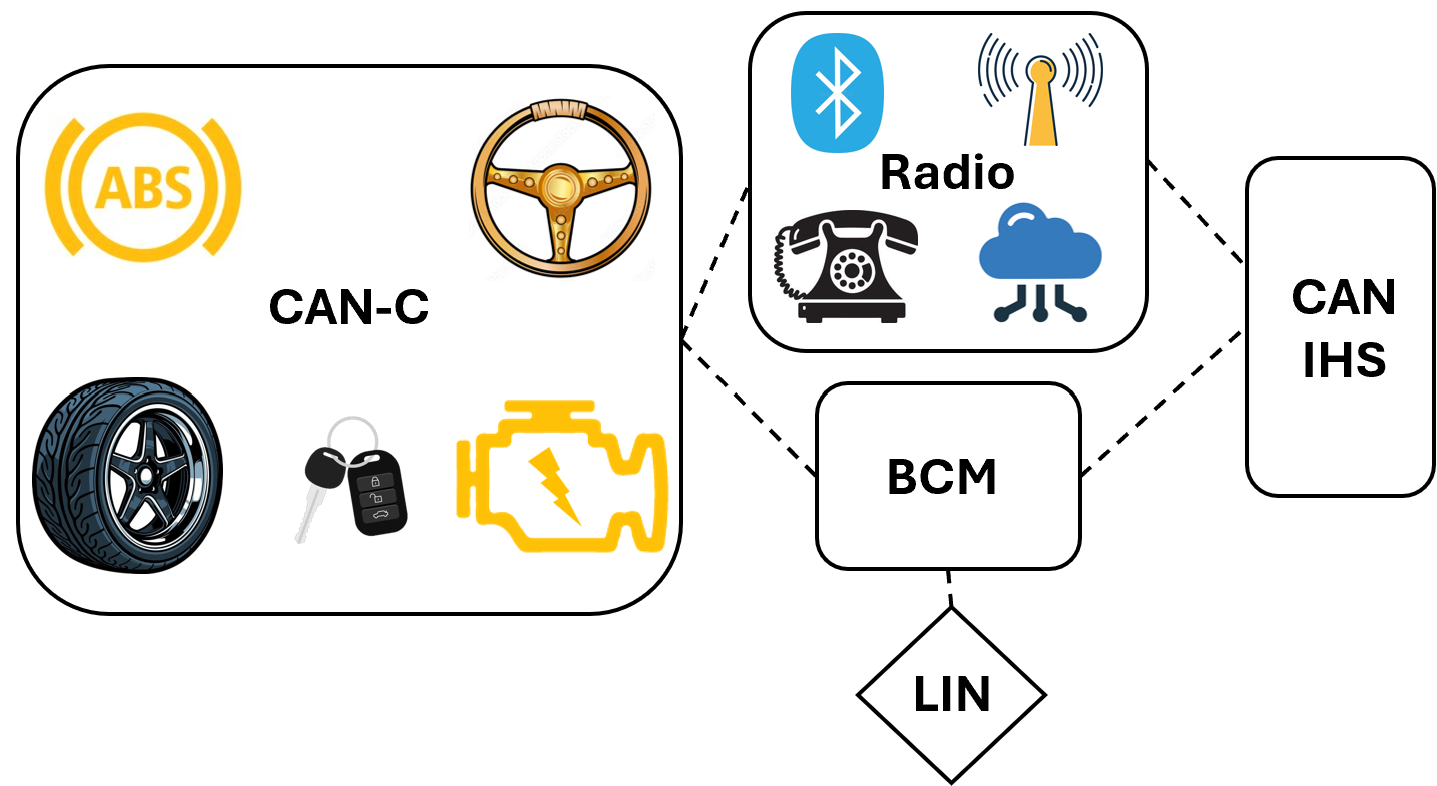}
\end{center}
\caption{Jeep Cherokee Network Architecture \cite{miller2015remote}.}
\label{fig:jeep_arch}
\end{figure}

A vehicle network interlinks all the Electronic Control Units (ECUs) within a vehicle, executing essential functions such as managing the engine and transmission. In autonomous vehicles, this network facilitates autonomous operations by enabling data sharing and the integration of advanced AI-driven sensors into the vehicle's systems. Semi- and fully autonomous vehicles are equipped with head units that connect to various vehicle networks, providing WiFi, Bluetooth, and cellular communications. The architecture of vehicle networks is highly varied, depending on the manufacturer, vehicle features, and pricing.\\
Figure \ref{fig:jeep_arch}{} illustrates the network architecture of a Jeep Cherokee (Level 2-3 Autonomy), which features a Controller Area Network Critical (CAN-C), a Controller Area Network Interior High Speed (CAN-IHS), and a Local Interconnect Network (LIN). The CAN-C network connects all critical Electronic Control Units (ECUs), such as the Anti-lock Braking System (ABS) unit and engine control systems, managing essential vehicle operations. The Body Control Module (BCM) interfaces with both the CAN-C and LIN, controlling sensors and actuators on the vehicle's body based on inputs from the CAN. Additionally, the CAN-C interconnects with the head unit through the CAN-IHS.

\begin{figure*}
\begin{center}
\includegraphics[clip,width=1.8\columnwidth]{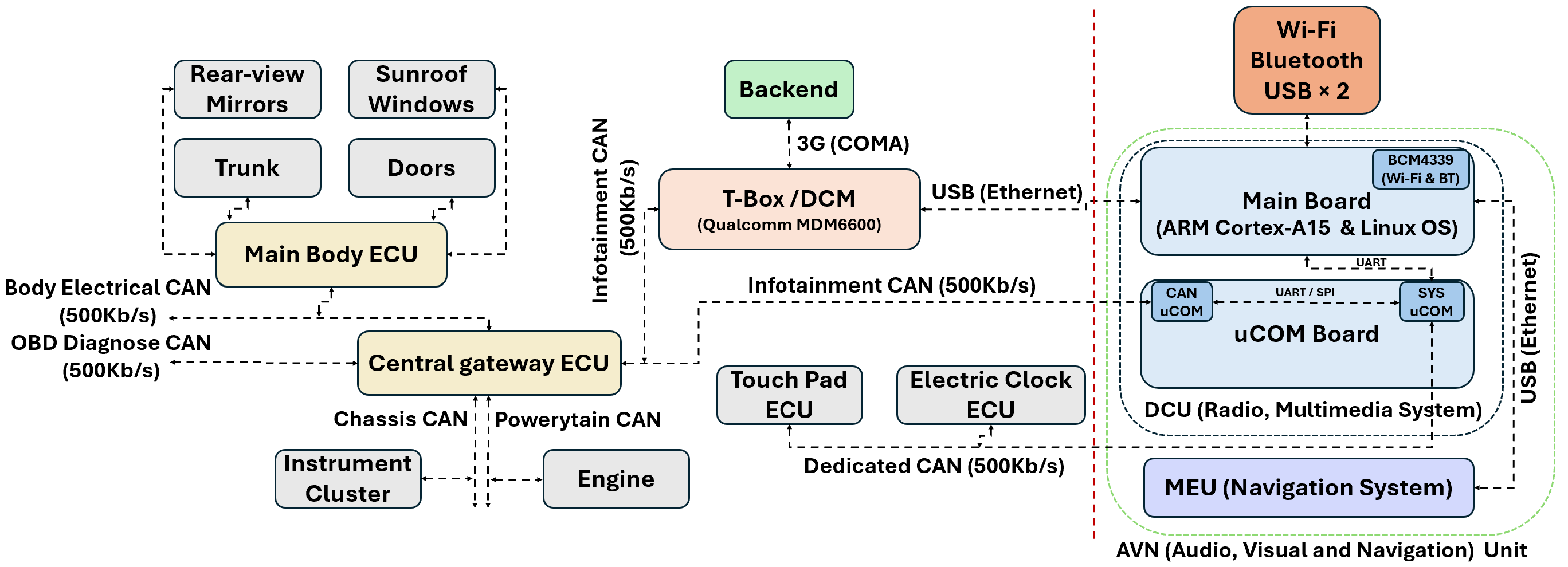}
\end{center}
\caption{Lexus Network Architecture \cite{tencent2020}.}
\label{fig:lexus_arch}
\end{figure*}

Figure \ref{fig:lexus_arch} presents the network architecture of a Lexus (Level 3-4 Autonomy), which differs from that of the Jeep Cherokee depicted in Figure \ref{fig:jeep_arch} The Lexus features a Central Gateway ECU that connects various Controller Area Network (CAN) networks. Specifically, the Lexus includes:
\begin{itemize}
    \item Body Electrical CAN: This network connects to the Main Body ECU, which manages all body functions of the vehicle via a Local Interconnect Network (LIN).
    \item Powertrain CAN: This network facilitates communication among all systems involved in the powertrain, encompassing components that generate energy for the vehicle, such as the engine and transmission.
    \item Chassis CAN: This network connects all body sensors and instrument clusters.
    \item Infotainment CAN: This network links entertainment units within the vehicle, connecting the head unit, clock, touchpad, etc., while also providing WiFi, Bluetooth, and cellular connectivity.
\end{itemize}

\begin{figure*}
\begin{center}
\includegraphics[clip,width=1.8\columnwidth]{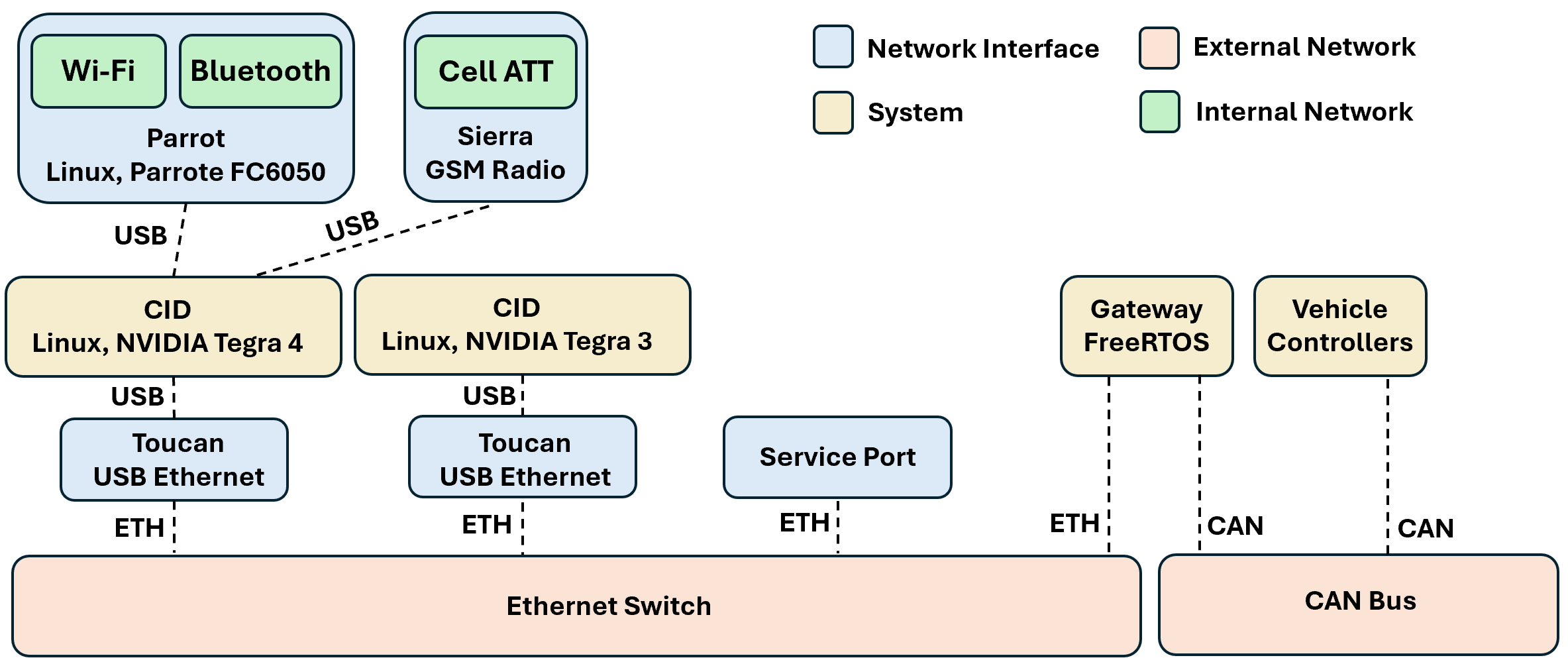}
\end{center}
\caption{Tesla Network Architecture \cite{mahaffey2015}.}
\label{fig:tesla_arch}
\end{figure*}

Figure \ref{fig:tesla_arch} illustrates the network architecture of a Tesla (Level 4-5 Autonomy), which differs significantly from the architectures of both the Lexus and the Jeep. Tesla employs Automotive Ethernet to interconnect all its Electronic Control Units (ECUs). This standard is specifically designed to meet the requirements of automotive applications, offering low susceptibility to electromagnetic and radio emissions, high bandwidth, and low latency. Additionally, Tesla utilizes a Linux-based NVIDIA head unit that manages all vehicle operations through a Controller Area Network (CAN) bus.

\begin{figure}
\begin{center}
\includegraphics[clip,width=0.7\columnwidth]{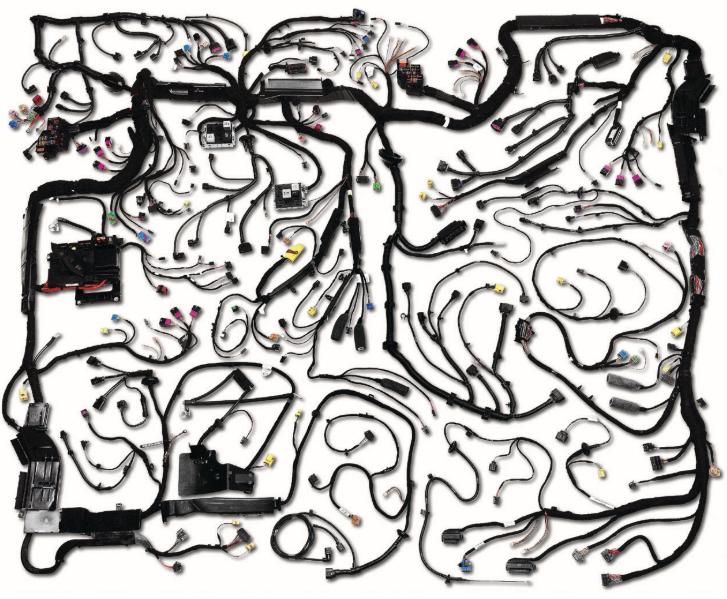}
\end{center}
\caption{Automation level 2 or 3 wiring harness \cite{ixia2014}.}
\label{fig:wiring}
\end{figure}

Figure \ref{fig:wiring} depicts the complexity of a vehicle's wiring harness at automation levels 2 or 3. The wiring harness has become increasingly intricate due to the integration of sensors and networks, reflecting the growing demands of modern automotive technology.

\subsection{Generic Autonomous Vehicle Architecture}
As discussed in previous sections, semi- and fully autonomous vehicles feature complex network architectures that differ based on the level of autonomy and price range. These architectures will only grow more intricate as vehicles become increasingly autonomous and interconnected. Furthermore, the earlier sections illustrate that even well-designed vehicles can harbor vulnerabilities, which attackers can exploit either in person or remotely to cause catastrophic failures. This underscores the critical need for a structured threat model to address the evolving security challenges in the automotive industry.\\

\begin{figure*}
\begin{center}
\includegraphics[clip,width=1.5\columnwidth]{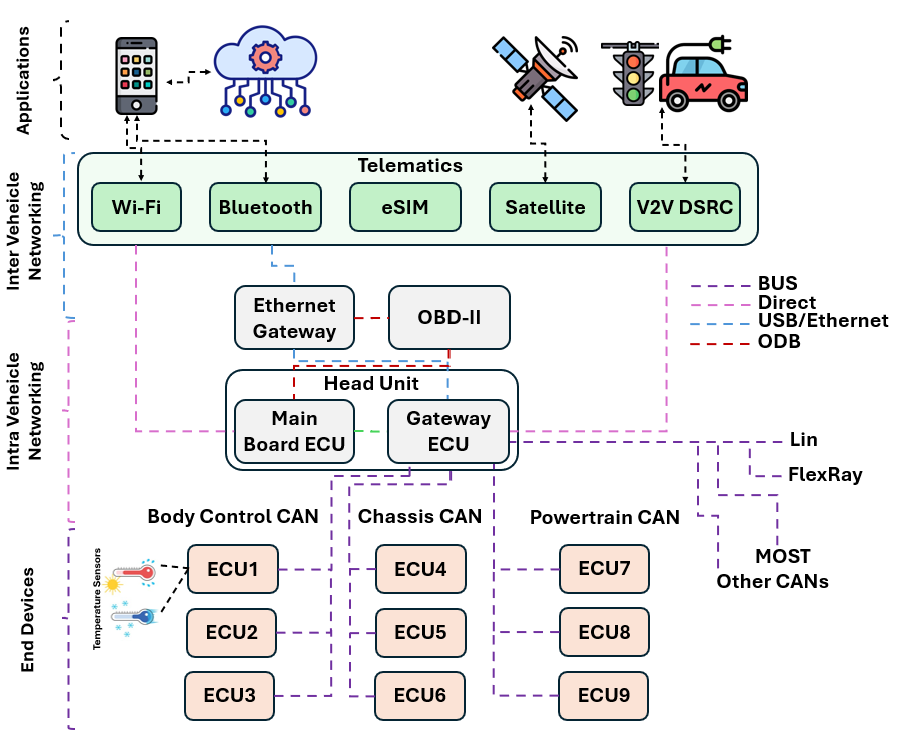}
\end{center}
\caption{Generic AV Architecture.}
\label{fig:generic_arch}
\end{figure*}

This section introduces a generic architecture for autonomous vehicles (AVs) (Figure \ref{fig:generic_arch}) to support threat modeling. This architecture is organized into four layers:
\begin{itemize}
    \item End Devices Layer: This layer consists of Electronic Control Units (ECUs) that manage and control various physical processes within the vehicle. These ECUs are physically connected to sensors, which monitor the vehicle, and actuators, which control its functions. The ECUs communicate over various intra-vehicle networks, such as CAN bus, LIN, MOST, and FlexRay.

    \item Intra-Vehicle Networking Layer: This layer facilitates communication between the vehicle’s ECUs to support management and control of operations. It consists of:

    \begin{itemize}
     
        \item Gateway ECU: Handles all communication with intra-vehicle networking buses (CAN, LIN, MOST, and FlexRay), ensuring message conformance to standards and preventing direct application access to the buses.
    
        \item Mainboard ECU: Manages the head unit and handles functions like navigation, display, and network management, using Universal Asynchronous Receiver Transmitter (UART) for sending and receiving CAN messages and commands.
    
        \item Telematics Gateway ECU: Manages data switching between the telematics modules and the head unit.
    
        \item OBD-II: Provides vehicle diagnostics and allows users to observe messages on the buses (such as the CAN bus). Through the OBD-II port, users can also send messages to control vehicle behavior.
    
        \item Vehicle Networks: Networks such as the CAN bus facilitate communication between ECUs controlling critical vehicle functions like the engine and transmission.
    \end{itemize}
    \item Inter-Vehicle Networks Layer: This layer connects the AV to external systems like smartphones and cloud services. It includes:
    \begin{itemize}
        \item eSIM: Enables communication over 3G, LTE, 4G, and 5G networks.
        \item Satellite Communications: Provides media broadcasts, real-time traffic updates, and navigation data.
        \item Bluetooth and WiFi: Allows connectivity with smartphones and hotspots, enabling media playback and Over-the-Air (OTA) updates. These networks are managed via the head unit.
    \end{itemize}
    \item Applications Layer: This layer consists of cloud services and other applications that provide critical functions for the vehicle's autonomous operations.
\end{itemize}

\section{Autonomous Vehicle cybersecurity Attacks}\label{sec:attacks}
The growing complexity and interconnectedness of modern vehicles make them increasingly vulnerable to cyberattacks. Legacy vehicles that relied on CAN and LIN networks were susceptible to attacks targeting their limited onboard sensors and ECUs. With the introduction of Vehicle-to-Everything (V2X) communications in semi- and fully autonomous vehicles, the range of potential attack vectors has expanded, exposing vehicles to remote cyberattacks in addition to traditional ones that target sensors and in-vehicle networks. Connected vehicles primarily use wireless communication (with exceptions such as electric vehicles during charging), which presents easy targets for attackers. By exploiting wireless channels, such as WiFi, Bluetooth, or cellular connections, attackers can compromise an ECU and escalate to more sophisticated vehicle cyberattacks. Researchers have already demonstrated successful breaches of vehicles via vulnerabilities in these wireless communication systems.\\
This section examines several real-world security incidents involving autonomous vehicles (AVs), providing insights into evolving threats and highlighting critical vulnerabilities across different AV systems. The analysis includes both historical and recent cases, including the Jeep Cherokee and Tesla Model S attacks, as well as recent vulnerabilities in Tesla Autopilot and Vehicle-to-Everything (V2X) communication systems.

\subsection{Common Cybersecurity Attack Vectors in AVs}
Autonomous vehicles (AVs) rely on interconnected systems, real-time communication, and sophisticated software, making them vulnerable to a wide range of cybersecurity attacks. One major category of attacks is through wireless communication channels such as Wi-Fi \cite{alipour2015wireless , satam2020wids, satam2017anomaly}, Bluetooth \cite{satam2018bluetooth, satam2020multi}, and cellular networks. Exploiting vulnerabilities in these channels allows attackers to execute remote exploits, gaining unauthorized control over vehicle functions. The widely publicized Jeep Cherokee hack by Miller and Valasek (2015) demonstrated how attackers could take control of braking, steering, and acceleration through cellular network vulnerabilities\cite{miller2015remote}. Similarly, Vehicle-to-Everything (V2X) communication, which enables AVs to interact with infrastructure (V2I) and other vehicles (V2V), is susceptible to message spoofing, replay attacks, and man-in-the-middle (MITM) attacks, which compromise situational awareness and lead to unsafe decisions\cite{petit2015cyberattacks}.
\\
Sensor spoofing and jamming represent another significant threat \cite{alshawi2020effective}. AVs depend on multiple sensors, including LiDAR, radar, cameras, and GPS, for perception and navigation. Attackers can exploit these sensors by injecting false signals or obstructing genuine signals. For instance, LiDAR spoofing can create phantom obstacles, causing the vehicle to brake suddenly or change lanes unnecessarily\cite{yan2016can}. GPS spoofing attacks can manipulate the vehicle's geolocation data, resulting in incorrect navigation or route deviations\cite{humphreys2012statement}. Additionally, radar and camera sensors are vulnerable to jamming or blinding attacks that disrupt data collection and processing, impairing the vehicle’s ability to detect objects accurately\cite{shoukry2013non}.
\\
The Controller Area Network (CAN) bus is a critical component in AVs, facilitating communication between Electronic Control Units (ECUs). Due to its lack of encryption and authentication mechanisms, the CAN bus is vulnerable to injection attacks that allow adversaries to send malicious commands to control essential functions like steering, braking, and acceleration. A well-documented example of this vulnerability is the ability to manipulate vehicle behavior through physical access to the CAN bus interface\cite{checkoway2011comprehensive}. These attacks highlight the need for enhanced security protocols within internal vehicle networks.
\\
Another vector is firmware and software vulnerabilities, particularly through Over-the-Air (OTA) updates. While OTA updates are crucial for maintaining and improving vehicle software, insecure update mechanisms can be exploited to install malicious firmware. If proper encryption, code-signing, and integrity verification are absent, attackers can compromise the update process to gain persistent control over vehicle functions\cite{halder2020secure}. This vulnerability underscores the importance of robust end-to-end encryption and authenticated update procedures to protect against malicious updates\cite{eiza2017driving}.

\subsection{Emerging Cybersecurity Threats}
As AV technology evolves, new cybersecurity threats are emerging, requiring proactive defenses. AI adversarial attacks pose a significant challenge to AV perception systems, which rely on machine learning models for object recognition and decision-making. In these attacks, adversaries introduce subtle perturbations to sensor inputs, such as images from cameras, causing the AV to misinterpret road signs or fail to detect obstacles\cite{eykholt2018robust}. For example, slight modifications to a stop sign can cause an AV to perceive it as a speed limit sign, potentially leading to dangerous driving behavior\cite{goodfellow2014explaining}. This vulnerability highlights the need for developing robust and resilient AI models capable of detecting and mitigating adversarial inputs in real-time.
\\
Supply chain vulnerabilities also present a growing concern. AVs rely on complex supply chains for hardware components and software systems, often sourced from multiple vendors across different regions. Compromises in the supply chain, such as inserting malicious code during manufacturing or embedding hardware backdoors, can undermine vehicle security\cite{wang2019understanding}. Recent incidents have demonstrated how vulnerabilities in third-party components can be exploited to bypass security controls, emphasizing the need for rigorous supply chain verification and secure development practices\cite{boyes2018industrial}.
\\
Cloud and edge computing vulnerabilities further complicate AV cybersecurity. AVs increasingly depend on cloud-based services and edge computing for data processing, storage, and decision-making. These infrastructures are susceptible to Distributed Denial-of-Service (DDoS) attacks, data breaches, and unauthorized access\cite{hou2016vehicular}. Disruptions to cloud services can lead to delays or failures in critical AV functions, affecting safety and performance. Ensuring secure communication channels, implementing redundancy, and adopting blockchain-based solutions can help mitigate these risks\cite{sharma2017distblocknet}.
\\
Finally, insider threats remain a persistent concern, where individuals with authorized access to AV systems exploit their privileges for malicious purposes. Such threats can arise from disgruntled employees, contractors, or malicious insiders who introduce vulnerabilities or exfiltrate sensitive data\cite{greitzer2014analysis}. Implementing zero-trust architectures, multi-factor authentication, and continuous monitoring can help mitigate insider threats by minimizing the potential for unauthorized actions\cite{schmittner2016using}.
\\
Addressing these diverse cybersecurity threats requires a multi-layered approach, combining traditional security measures with emerging technologies like blockchain, AI-driven threat detection, and comprehensive threat modeling frameworks such as STRIDE and MITRE ATT\&CK\cite{hussain2019autonomous, khan2020cyber}. Collaboration between AV developers, cybersecurity researchers, and policymakers is essential to developing robust security strategies that keep pace with evolving threats and ensure the safety and reliability of autonomous vehicles.
\subsection{Mitigation Strategies}
\textbf{Securing Wireless Communications:}
\begin{itemize}
    \item Encryption and Authentication: Use end-to-end encryption and mutual authentication protocols like Transport Layer Security (TLS) for all wireless communication \cite{hussain2019autonomous}.
    \item Network Isolation: Segregating infotainment systems from critical vehicle controls minimizes the impact of wireless exploits \cite{miller2015remote}.
\end{itemize}

\textbf{Enhancing Sensor and AI Resilience:} 
\begin{itemize}
    \item Dynamic Watermarking: Embed unique identifiers in sensor signals to detect tampering \cite{shangguan2022dynamic}.
    \item Adversarial Training: Expose AI models to adversarial scenarios during training to increase robustness against manipulated inputs \cite{girdhar2023cybersecurity}.
    \item Cross-Sensor Validation: Integrate data from multiple sensors (e.g., radar, LiDAR, cameras) to cross-check and validate environmental information \cite{khayyam2020artificial}.
    \item Redundancy and Validation: Use diverse sensors to ensure comprehensive cross-validation of sensor data \cite{saeed2023review}.
\end{itemize}

\textbf{Securing V2X Communication:} 
\begin{itemize}
    \item Blockchain Technology: Utilize decentralized ledgers to validate V2X messages and prevent spoofing attacks \cite{kukkala2022roadmap}.
    \item Public-Key Infrastructure (PKI): Deploy PKI to ensure the authenticity and integrity of transmitted messages \cite{alotaibi2018secure}.
    \item Enhanced Authentication: Implement multi-factor authentication mechanisms for V2X endpoints \cite{folan2023cybersecurity}.
\end{itemize}

\textbf{Protecting Firmware and Software:}
\begin{itemize}
    \item Secure OTA Updates: Use code-signing mechanisms to verify the integrity of over-the-air updates \cite{miller2015remote}.
    \item Static and Dynamic Analysis: Regularly test AV software for vulnerabilities using automated analysis tools \cite{potteiger2016software}.
    \item Secure Boot Mechanisms: Ensure only authenticated firmware can be executed during system boot-up \cite{eiza2017driving}.
    \item Runtime Monitoring: Continuously validate system integrity during operation to detect malicious activity \cite{khan2020cyber}.
\end{itemize}

 \textbf{Hardening In-Vehicle Networks:} 
 \begin{itemize}
     \item Intrusion Detection Systems (IDS): Deploy IDS to monitor network traffic and identify anomalous activity \cite{hussain2019autonomous}.
     \item Message Authentication Codes (MACs): Use MACs to verify the authenticity of messages transmitted over the CAN bus \cite{kukkala2022roadmap}.
 \end{itemize}

\textbf{Human Factors and Trust Repair:}
\begin{itemize}
    \item Transparent Communication: Provide detailed updates to users following an attack, explaining the mitigation measures to rebuild trust \cite{lim2024impact}.
    \item Driver-AI Collaboration Models: Design AV systems that adaptively recalibrate trust levels post-attack through user interaction and feedback mechanisms \cite{linkov2019human}.
\end{itemize}

\subsection{Case Studies}
\subsubsection{Case study 1: Jeep Cherokee Remote Exploit (2015)}
\begin{figure*}
\begin{center}
\includegraphics[clip,width=1.8\columnwidth]{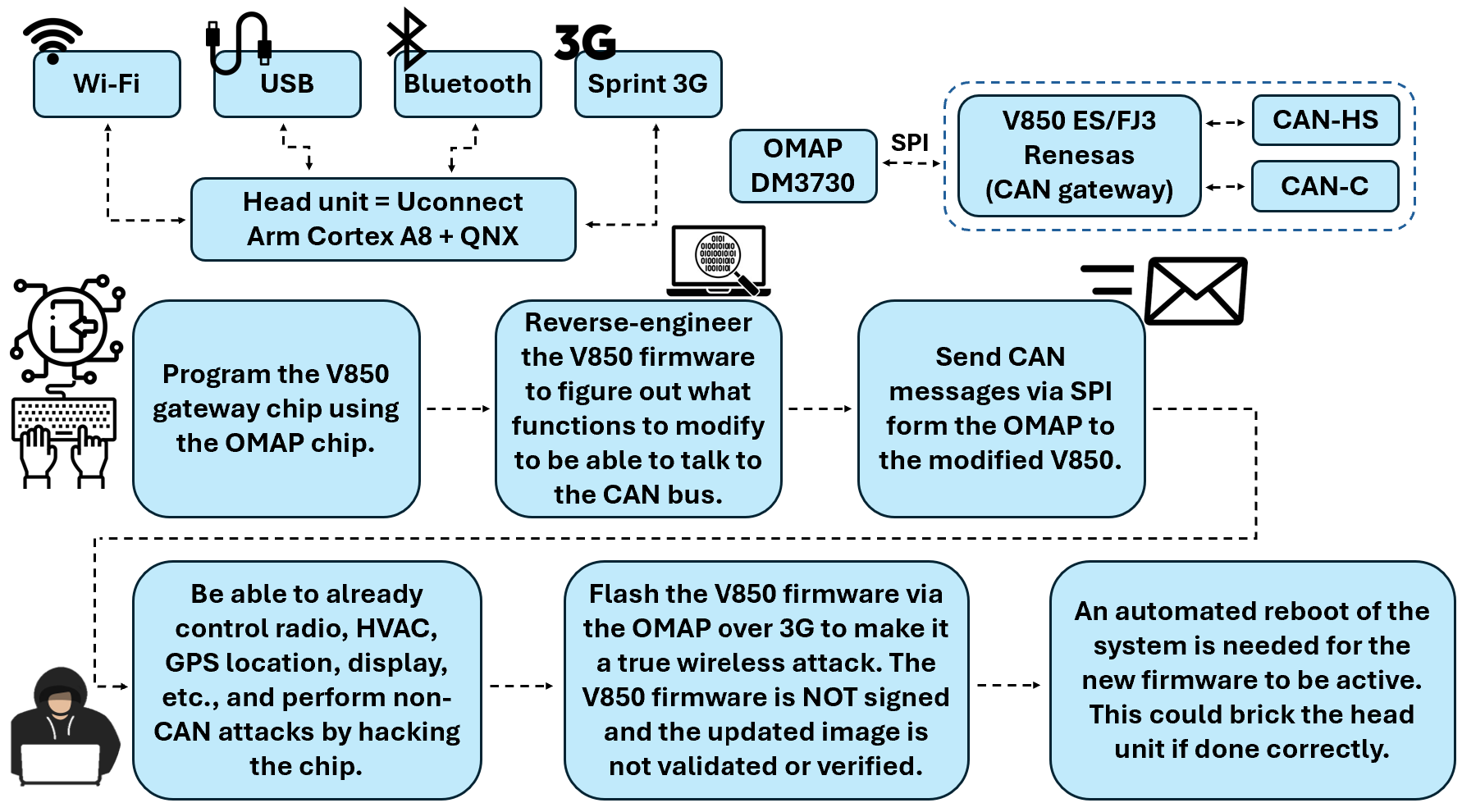}
\end{center}
\caption{Jeep Cherokee remote attack chain \cite{miller2015remote}.}
\label{fig:jeep__attack_chain}
\end{figure*}

In 2015, cybersecurity researchers exploited a vulnerability in the Jeep Cherokee’s UConnect infotainment system, remotely gaining control over critical functions, such as braking and steering. By using an exposed port to access the Controller Area Network (CAN) bus, the attackers were able to inject malicious CAN messages that disabled essential safety features. This attack underscored the importance of secure communication channels and authentication in vehicular networks \cite{miller2015remote}.\\
Figure \ref{fig:jeep__attack_chain} illustrates the attack chain. The Jeep Cherokee's head unit, powered by an Arm Cortex A8 processor with UConnect, provides connectivity through WiFi, Bluetooth, USB, and cellular networks (Sprint 3G), linking the vehicle to external systems. The researchers demonstrated that, with physical access, an attacker could compromise and reprogram the head unit via USB.\\
Additionally, the researchers targeted the head unit's WiFi and Bluetooth connections by cracking the password, granting unauthorized access to the vehicle’s network when within visual range. They also discovered that Jeep UConnect, through Sprint's 3G network, used port 6667 to run D-Bus message daemons, allowing the transmission of unauthenticated commands through Telnet and enabling command-line injection of supported Remote Procedure Call (RPC) methods. Sprint’s 3G network further allowed Sprint devices to communicate over vast distances, creating a nationwide Wide Area Network (WAN). This WAN setup, in conjunction with the exposed port 6667, enabled the creation of a network worm capable of infecting all vulnerable Jeep Cherokees with exposed port 6667.\\

The researchers developed a worm targeting the Jeep Cherokee's head unit, which used a Renesas V850 microprocessor, disguising it as a legitimate firmware update. Through this worm, they were able to inject arbitrary CAN messages onto the vehicle's CAN bus, gaining access to vendor-specific CAN messages used on the CAN-C network (Figure \ref{fig:jeep_arch}). This enabled them to remotely control critical vehicle functions, such as killing the engine, disabling the brakes, and disengaging the steering wheel.

\textbf{Response to Jeep Cherokee Exploit:} After the Jeep Cherokee exploit, automakers took steps to isolate critical vehicle control functions from infotainment systems and other non-essential communication channels. Security updates included closing exposed network ports and implementing strict access controls to the CAN bus. These measures aimed to prevent external manipulation of essential vehicle controls through enhanced network segmentation and control channel isolation \cite{miller2015remote}.

\subsubsection{Case study 2: Tesla Model S Exploit by Tencent Keen Security Lab (2016)}
\begin{figure*}
\begin{center}
\includegraphics[clip,width=1.8\columnwidth]{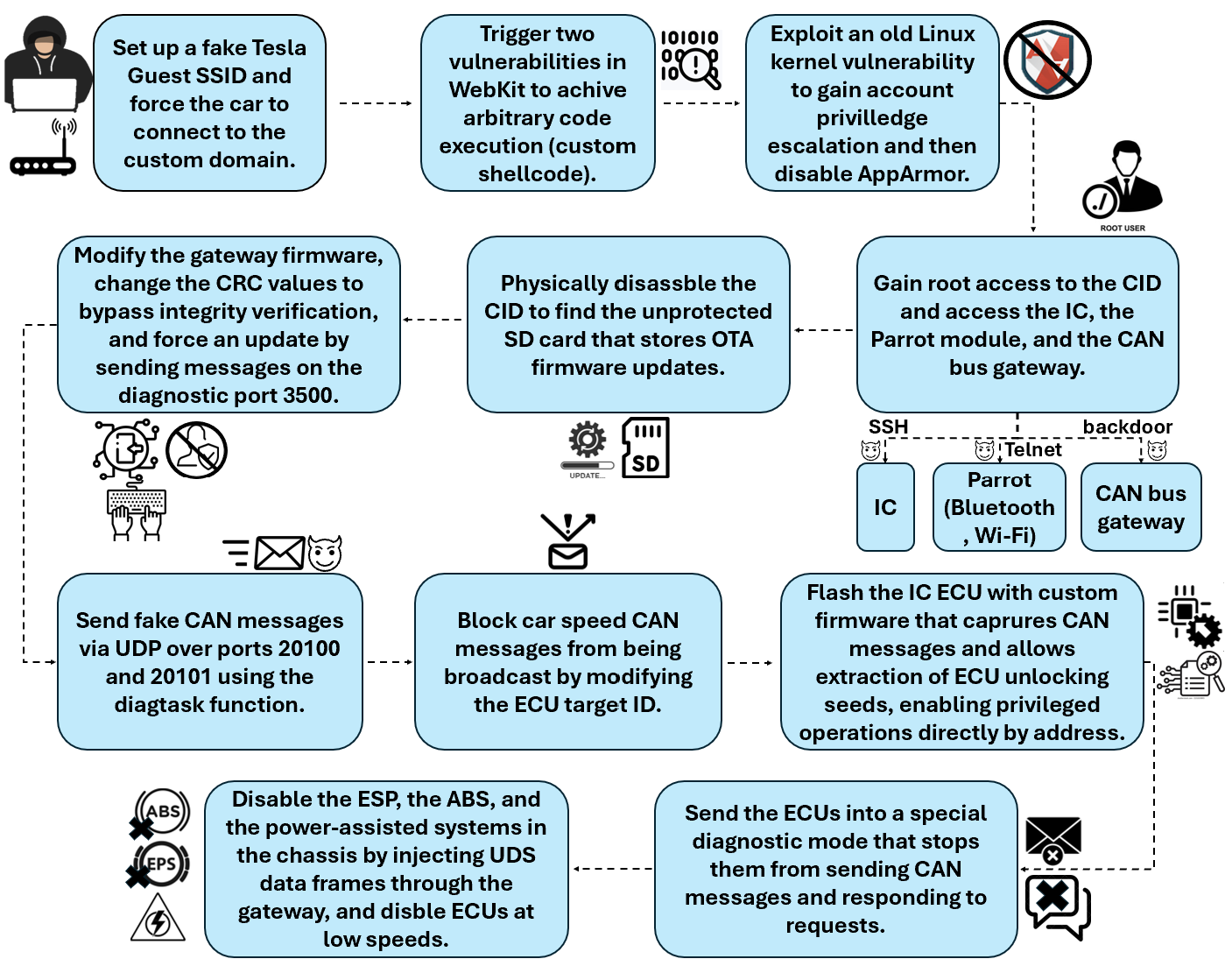}
\end{center}
\caption{Tesla remote attack chain \cite{nie_blackhat}.}
\label{fig:tesla__attack_chain}
\end{figure*}

In 2016, researchers at Tencent Keen Security Lab compromised a Tesla Model S by exploiting vulnerabilities in its onboard WiFi connection and browser. By manipulating the web browser’s access, they gained control over the CAN bus, enabling them to execute commands that affected steering and braking. This attack demonstrated the risk of unsecured wireless communication channels and the potential for remote compromise if external access is not properly managed. They discovered that all Tesla vehicles are configured to automatically connect to the SSID "Tesla Guest," a WiFi hotspot used by Tesla superchargers and body shops. Additionally, the researchers found that the Tesla Model S uses a WebKit-based web browser engine running on a Linux system. The attackers set up a rogue access point named "Tesla Guest," causing the car to automatically connect to their controlled network.\\
Once connected, the attackers exploited a known Linux kernel vulnerability, CV2013-6282 \cite{nie_blackhat}, to escalate privileges within the WebKit browser. By further targeting vulnerabilities in WebKit, they were able to gain remote shell access and execute custom shellcode. With this elevated WebKit access, the researchers gained control of Tesla's Central Information Display (CID), which provided access to the Instrument Cluster (IC), as well as the Bluetooth and WiFi modules via Telnet, and the CAN bus through a custom backdoor.\\
To create the CAN bus backdoor, the researchers observed that the Tesla Gateway (Figure \ref{fig:tesla_arch}) is enabled for over-the-air (OTA) updates and uses port 3500 for diagnostics, while ports 20100 and 20101 are used for sending CAN messages. By extracting the OTA firmware package from the CID through their privileged access, they modified its Cyclic Redundancy Check (CRC) to bypass the integrity verification. The compromised firmware was then installed on the Gateway via the diagnostics port 3500, enabling the researchers to send malicious CAN messages over User Datagram Protocol (UDP). Figure \ref{fig:tesla__attack_chain} outlines the entire attack chain.
Additionally, Tesla ECUs have a special diagnostic mode that halts the ECU from sending CAN messages and responding to diagnostic requests. With control over the compromised Gateway, the researchers were able to force any ECU into this diagnostic mode, effectively disabling critical vehicle systems. This allowed them to deactivate essential safety features such as the Electronic Stability Program (ESP), the Anti-lock Braking System (ABS), and the power-assisted chassis, rendering these systems non-functional.\\
\textbf{Tesla’s Mitigation for Model S Wireless Exploit:} In response to the wireless exploit, Tesla deployed over-the-air (OTA) updates to secure the WiFi and browser components of its vehicles, limiting external access to critical systems. Tesla also introduced stricter network access controls and improved the authentication protocols to prevent unauthorized access to CAN bus commands, reinforcing protection against remote attacks on the vehicle’s control systems 

\subsubsection{Case study 3:Tesla Autopilot Adversarial Attack (2021)}
In 2021, researchers demonstrated how Tesla’s Autopilot could be misled through adversarial attacks on the vehicle’s visual recognition systems. By subtly altering road signs or lane markings, attackers could influence the Autopilot system to misinterpret inputs, causing unsafe lane changes or sudden stops. This exploit illustrated the vulnerability of AVs to image-based manipulation and the need for robust AI training against adversarial inputs \cite{girdhar2023cybersecurity}\\
\textbf{Countermeasures for Adversarial Attacks on Tesla Autopilot:} To counteract adversarial threats, Tesla began enhancing its AI models through adversarial training, exposing the system to manipulated inputs to make it more resilient to subtle image-based distortions. Tesla also implemented cross-sensor validation, where visual data is cross-referenced with radar or LiDAR inputs, adding an additional layer of verification before actions like lane changes or braking are executed \cite{girdhar2023cybersecurity}.

\subsubsection{Case study 4: V2X Communication Spoofing}
Vehicle-to-Everything (V2X) communication systems facilitate data exchange between AVs and infrastructure (such as traffic lights and road signs), improving traffic management and safety. However, these systems are vulnerable to spoofing attacks, where attackers inject false information into the network. In a V2X spoofing scenario, a vehicle could receive a false warning about obstacles or phantom vehicles, prompting unsafe braking or swerving. This attack highlights the critical need for V2X message authentication to ensure data integrity and secure communication \cite{kukkala2022roadmap}.\\
\textbf{Defenses Against V2X Communication Spoofing:} To mitigate V2X communication attacks, researchers are exploring the use of blockchain and public-key infrastructure (PKI) for secure message authentication. Blockchain can offer a tamper-resistant ledger to validate message authenticity, while PKI can secure V2X data exchanges, ensuring that only legitimate data influences AV responses. These cryptographic solutions aim to safeguard V2X networks from spoofing and tampering, allowing AVs to communicate securely with infrastructure and other vehicles \cite{khayyam2020artificial}.

\section{Overview of Threat Modelling}\label{sec:modelling}
Threat modeling is a structured process to assess potential actions attackers might take, identify security vulnerabilities and threats within a system or its components, and determine which assets could be exploited. This process provides valuable insights for protecting the target system or sub-system against attacks and helps develop realistic and effective security requirements \cite{hamad2020savta}.

\subsection{Threat Modeling Methods}
Various methods of conducting threat modeling on a system emphasize different aspects such as assets, attackers, software, or system vulnerabilities. The following are some of the key threat modeling approaches that can be applied to a system:

\subsubsection{Attack-Centric Threat Modeling}
Attack-centric modeling focuses on understanding attackers' goals, motivations, and required skills to target a specific system \cite{hamad2020savta}. By analyzing the attacker’s perspective, we can gain insight into their objectives, identify the most vulnerable assets, and design effective mitigation strategies to counter their actions. A notable example of an attack-centric threat model is the Intel Threat Agent Library (TAL) \cite{casey2007threat}. TAL standardizes a set of agents that pose threats to IT systems and classifies them based on eight key questions: 1) the outcome of the threat, 2) the resources available to the attacker, 3) whether the attacker intends to cause harm, 4) the limitations that could hinder the attack, 5) the attacker's objective, 6) the identity of the threat, 7) the skills required to carry out the attack, and 8) the attacker's goals.\\
Another example of threat modeling is Intel's Threat Agent Risk Assessment (TARA) \cite{casey2007threat}. TARA helps identify all potential attacks on a system and determines the most likely attack scenarios based on the Intel Threat Agent Library (TAL). This method provides a focused approach to assessing risks and prioritizing mitigation strategies by evaluating the probability of specific threats.

\subsubsection{Asset-Centric Threat Modeling}
Asset-centric threat modeling focuses on the system's assets and how to protect them from various attacks \cite{casey2007threat}. This approach involves identifying attack surfaces, assessing vulnerabilities, and pinpointing critical assets whose compromise would cause the most harm to the system.\\
An example of asset-centric threat modeling is the Operationally Critical Threat, Asset, and Vulnerability Evaluation (OCTAVE) \cite{alberts2003introduction}, which includes the following four phases: 1) defining risk measurement criteria, 2) profiling critical assets, 3) identifying threats to those assets, and 4) conducting risk analysis and developing mitigation strategies.\\
Another example is the Threat, Vulnerability, and Risk Analysis (TVRA) \cite{woodworth2015protection}. The TVRA model outlines the security requirements and objectives for each asset in the system, performs vulnerability scanning, and assesses the likelihood of potential threats.

\subsubsection{Threat-Centric Modeling and Vulnerability}
In threat-centric and vulnerability modeling, high-priority vulnerabilities are identified, and mitigation techniques are developed to address them \cite{casey2007threat, potteiger2016software}. An example of this approach is Skybox Threat-Centric Vulnerability Management (TCVM), which generates a comprehensive list of vulnerabilities and then prioritizes them based on their potential impact and likelihood of exploitation. This method helps focus mitigation efforts on the most critical vulnerabilities.

\subsubsection{Software-Centric Threat Modeling}
Software-centric threat modeling, also known as design-centric threat modeling, focuses on the software components of a system during the design phase. This approach helps identify potential threats to each component \cite{casey2007threat}.\\
A well-known example of software-centric threat modeling is Microsoft's STRIDE model, which stands for Spoofing, Tampering, Repudiation, Information Disclosure, Denial of Service, and Elevation of Privilege. STRIDE is used to identify and address potential threats across these categories, making it particularly useful for systems like autonomous vehicles.

\subsection{Attack Trees}
An attack tree is a visual representation of an attack, structured like a tree \cite{schneier1999features}. The root node represents the attacker's primary goal, while the intermediate nodes, or sub-goals, illustrate the various stages of the attack. For the attack to succeed, all sub-goals in a particular branch must be achieved. The leaf nodes represent atomic attacks—individual actions required to complete an attack. Each attack scenario is composed of the minimum set of nodes. Attack trees are useful for representing multiple possible attack paths on a system, aiding in risk analysis, and identifying potential attack surfaces that an attacker could exploit.

\subsection{MITRE ATT\&CK}
MITRE ATT\&CK (Adversarial Tactics, Techniques, and Common Knowledge) is a publicly accessible knowledge base of adversary tactics and techniques, derived from real-world observations of cyberattacks \cite{kwon2020cyber}. It serves as a foundation for developing threat models and methodologies across various sectors, including government, private companies, and the cybersecurity product and service community. MITRE ATT\&CK catalogs strategies that attackers use to execute successful attacks and lists the techniques that can implement these strategies. **Tactics** represent what an attacker aims to achieve, while **techniques** describe how the attacker accomplishes these objectives.\\

MITRE ATT\&CK consists of three primary iterations:
\begin{enumerate}
    \item \textbf{ATT\&CK for Enterprise} focuses on adversarial behavior across Windows, macOS, Linux, and cloud environments.
    \item \textbf{ATT\&CK for Mobile} targets adversarial behavior on Android and iOS operating systems.
    \item \textbf{ATT\&CK for ICS} addresses adversarial behavior in Industrial Control Systems (ICS) networks.
\end{enumerate}

In ATT\&CK for ICS, approximately 300 attack tactics are organized in the knowledge base and categorized by threats, attack processes, and techniques. The relationship between tactics and techniques is visually represented through the ATT\&CK matrix \cite{alexander2020mitre}.

\section{STRIDE \& DREAD Threat Model for Autonomous Vehicles Architecture}\label{sec:strideanddread}
A threat model provides a structured approach to analyzing a system by examining each threat and identifying appropriate mitigation techniques. For this threat model, we use both the STRIDE and DREAD models to perform threat analysis and develop countermeasures.

\subsection{STRIDE Model}
STRIDE is a widely used threat modeling framework introduced in 1999 and popularized by Microsoft in 2002. The acronym STRIDE stands for six categories of threats that the model focuses on to evaluate a system's security:
\begin{itemize}
    \item S: Spoofing - Attacks where an attacker impersonates someone or something else, compromising system authentication.
    \item T: Tampering - Attacks where an attacker alters or modifies data, either on disks, networks, or in memory, compromising data integrity.
    \item R: Repudiation - Attacks where entities deny performing actions, often due to inadequate logging, making it difficult to trace actions.
    \item I: Information Disclosure - Attacks where unauthorized access to sensitive information occurs, breaching system confidentiality.
    \item D: Denial of Service - Attacks where an attacker exhausts system resources (computing or networking), reducing its functionality.
    \item E: Elevation of Privilege - Attacks where an attacker gains unauthorized access to higher-level functionalities or capabilities within the system, compromising authorization.
\end{itemize}

\subsection{DREAD Model}
The STRIDE model helps identify various threats targeting a system. Once these threats are identified, mitigating them involves several steps: 1) restructuring the system, 2) developing Intrusion Detection and Prevention Systems, 3) investing time, and 4) allocating money. Prioritizing threats is crucial to address the most damaging ones first and the less impactful ones later.\\
The DREAD model aids in prioritizing these threats by evaluating their impact. The acronym DREAD stands for:
\begin{itemize}
    \item Damage Potential: The extent of damage an attack could inflict on the asset.
    \item Reproducibility: How easily the attack can be replicated.
    \item Exploitability: How simple it is to launch the attack.
     \item Affected Users: The number of users that could be impacted by the attack.
    \item Discoverability: How easy it is to identify the exploitable weakness.
\end{itemize}
By assessing these factors, the DREAD model helps prioritize threats from most critical to least significant.\\

In the threat modeling process, each threat is categorized using the STRIDE model to determine the type of attack. This helps in answering the following questions:
\begin{itemize}
    \item Does this attack involve spoofing?
    \item Does this attack involve tampering of information?
    \item Does this attack allow repudiation?
    \item Does this attack target information disclosure?
    \item Does this attack cause a Denial of Service?
    \item Does this attack cause elevation of privilege?
\end{itemize}
Next,the impact of each threat is assessed using the DREAD model, where each threat is rated in three impact categories: High impact (3 points), Medium impact (2 points), and Low impact (1 point). The overall risk rating for a threat is calculated by summing the DREAD impacts.\\
The risk levels are categorized as follows:
\begin{itemize}
    \item An overall risk rating of 12-15 is considered high.
    \item A rating of 8-11 is considered medium.
    \item A rating of 5-7 is considered low.
\end{itemize}
Table \ref{tab:stride_dread} illustrates the threat model used in this process.

\section{Comparative Analysis of Threat Modeling Frameworks for Autonomous Vehicles}\label{sec:analysis}
Securing autonomous vehicles (AVs) requires robust threat modeling frameworks that can systematically identify and address various cybersecurity risks across complex, interconnected systems. This section provides a comparative analysis of three prominent threat modeling frameworks—STRIDE, DREAD, and MITRE ATT\&CK—based on their applicability, strengths, and limitations in the context of AV security.\\
Each framework brings unique benefits and limitations to AV cybersecurity. STRIDE is valuable for its structured threat categories, making it suitable for generalized threat identification; DREAD’s prioritization framework is ideal for assessing threat impact on real-time AV operations; and MITRE ATT\&CK offers a high level of detail for modeling specific attack vectors but can be resource-intensive. Combining elements from each model may yield the most robust threat modeling approach for AVs, enabling comprehensive security coverage from both a high-level and detailed perspective 
\subsection{STRIDE Model}
The STRIDE framework, initially developed by Microsoft, categorizes threats into six types: Spoofing, Tampering, Repudiation, Information Disclosure, Denial of Service, and Elevation of Privilege. STRIDE is particularly useful for identifying and organizing threats across AV components, as it offers a structured approach to assess and mitigate risks in individual system areas \cite{girdhar2023cybersecurity}.\\
STRIDE is highly adaptable to the layered architecture of AVs, where it can be used to isolate specific threats to vehicle sensors, control units, and communication protocols. By systematically covering different threat categories, STRIDE is effective in addressing diverse security aspects of AVs. However, STRIDE’s broad threat categories can sometimes lack the depth required for sophisticated attacks specific to autonomous systems, such as adversarial AI-based threats targeting sensor data. As a result, it may require supplementation with other models for complex threat scenarios \cite{aldhyani2022attacks}.

\subsection{DREAD Model}
DREAD evaluates threats based on Damage, Reproducibility, Exploitability, Affected Users, and Discoverability. This framework is effective for prioritizing threats, as it assigns numerical scores to each factor, enabling AV security teams to rank and focus on the most critical vulnerabilities \cite{hataba2022security}.\\
DREAD’s quantifiable scoring system is beneficial for AV security, where threat prioritization is crucial due to resource constraints. By evaluating the impact and ease of exploitation, DREAD helps determine where defenses should be strongest—especially for real-time systems in AVs, such as sensors and communication protocols. However, DREAD’s emphasis on impact factors might overlook the broader context of cyber-physical interactions in AVs. For example, it may be less effective for assessing complex, layered attacks targeting multiple AV subsystems simultaneously \cite{kukkala2022roadmap}.

\subsection{MITRE ATT\&CK Framework}
The MITRE ATT\&CK framework offers a comprehensive matrix of adversary tactics and techniques, derived from real-world observations of cyberattacks. This matrix-based approach is particularly valuable for AVs, as it allows for the modeling of sophisticated attack sequences and provides a library of specific threat techniques applicable across multiple AV components \cite{kyrkou2020towards}. MITRE ATT\&CK’s detailed threat taxonomy enables AV security teams to model realistic attack vectors specific to AVs, such as camera or sensor spoofing. The framework’s extensiveness allows for comprehensive threat mapping and the development of tailored countermeasures against specific attack tactics. However, due to its complexity, the MITRE ATT\&CK matrix can be resource-intensive to implement, requiring significant expertise to tailor its broad tactics and techniques to the AV domain. This complexity can be a challenge for organizations without dedicated cybersecurity resources \cite{girdhar2023cybersecurity}.

\onecolumn

\begin{table*}[t]
\caption{STRIDE \& DREAD Threat model for generic Autonomous Vehicle Architecture.}
\label{tab:stride_dread}

\centering
\begin{tabular}{|c|c|c|c|cccccc|}
\hline
\rowcolor[HTML]{C0C0C0} 
\cellcolor[HTML]{C0C0C0}                                                                                            & \cellcolor[HTML]{C0C0C0}                                                                                                             & \cellcolor[HTML]{C0C0C0}                                                                                                                   & \cellcolor[HTML]{C0C0C0}                                  & \multicolumn{6}{c|}{\cellcolor[HTML]{C0C0C0}\textbf{Threat Impact}}                                                                                                                                                                                                                                                                   \\ \cline{5-10} 
\rowcolor[HTML]{C0C0C0} 
\multirow{-2}{*}{\cellcolor[HTML]{C0C0C0}\textbf{\begin{tabular}[c]{@{}c@{}}AV Architecture \\ Layer\end{tabular}}} & \multirow{-2}{*}{\cellcolor[HTML]{C0C0C0}\textbf{Asset}}                                                                             & \multirow{-2}{*}{\cellcolor[HTML]{C0C0C0}\textbf{Attack Surface}}                                                                          & \multirow{-2}{*}{\cellcolor[HTML]{C0C0C0}\textbf{STRIDE}} & \multicolumn{1}{c|}{\cellcolor[HTML]{C0C0C0}\textbf{D}} & \multicolumn{1}{c|}{\cellcolor[HTML]{C0C0C0}\textbf{R}} & \multicolumn{1}{c|}{\cellcolor[HTML]{C0C0C0}\textbf{E}} & \multicolumn{1}{c|}{\cellcolor[HTML]{C0C0C0}\textbf{A}} & \multicolumn{1}{c|}{\cellcolor[HTML]{C0C0C0}\textbf{D}} & \textbf{Rating}                     \\ \hline
                                                                                                                    &                                                                                                                                      &                                                                                                                                            & T                                                         & \multicolumn{1}{c|}{3}                                  & \multicolumn{1}{c|}{3}                                  & \multicolumn{1}{c|}{3}                                  & \multicolumn{1}{c|}{1}                                  & \multicolumn{1}{c|}{3}                                  & \cellcolor[HTML]{FD6864}\textbf{13} \\ \cline{4-10} 
                                                                                                                    &                                                                                                                                      & \multirow{-2}{*}{\begin{tabular}[c]{@{}c@{}}Physical damage to the \\ Sensors or ECU\end{tabular}}                                         & D                                                         & \multicolumn{1}{c|}{3}                                  & \multicolumn{1}{c|}{3}                                  & \multicolumn{1}{c|}{3}                                  & \multicolumn{1}{c|}{3}                                  & \multicolumn{1}{c|}{3}                                  & \cellcolor[HTML]{FD6864}\textbf{15} \\ \cline{3-10} 
                                                                                                                    &                                                                                                                                      & \cellcolor[HTML]{EFEFEF}                                                                                                                   & \cellcolor[HTML]{EFEFEF}S                                 & \multicolumn{1}{c|}{\cellcolor[HTML]{EFEFEF}3}          & \multicolumn{1}{c|}{\cellcolor[HTML]{EFEFEF}1}          & \multicolumn{1}{c|}{\cellcolor[HTML]{EFEFEF}1}          & \multicolumn{1}{c|}{\cellcolor[HTML]{EFEFEF}1}          & \multicolumn{1}{c|}{\cellcolor[HTML]{EFEFEF}1}          & \cellcolor[HTML]{CBFE9B}\textbf{7}  \\ \cline{4-10} 
                                                                                                                    &                                                                                                                                      & \cellcolor[HTML]{EFEFEF}                                                                                                                   & \cellcolor[HTML]{EFEFEF}T                                 & \multicolumn{1}{c|}{\cellcolor[HTML]{EFEFEF}3}          & \multicolumn{1}{c|}{\cellcolor[HTML]{EFEFEF}2}          & \multicolumn{1}{c|}{\cellcolor[HTML]{EFEFEF}2}          & \multicolumn{1}{c|}{\cellcolor[HTML]{EFEFEF}1}          & \multicolumn{1}{c|}{\cellcolor[HTML]{EFEFEF}1}          & \cellcolor[HTML]{FFFC9E}\textbf{9}  \\ \cline{4-10} 
                                                                                                                    &                                                                                                                                      & \cellcolor[HTML]{EFEFEF}                                                                                                                   & \cellcolor[HTML]{EFEFEF}R                                 & \multicolumn{1}{c|}{\cellcolor[HTML]{EFEFEF}2}          & \multicolumn{1}{c|}{\cellcolor[HTML]{EFEFEF}1}          & \multicolumn{1}{c|}{\cellcolor[HTML]{EFEFEF}1}          & \multicolumn{1}{c|}{\cellcolor[HTML]{EFEFEF}1}          & \multicolumn{1}{c|}{\cellcolor[HTML]{EFEFEF}1}          & \cellcolor[HTML]{CBFE9B}\textbf{6}  \\ \cline{4-10} 
                                                                                                                    &                                                                                                                                      & \cellcolor[HTML]{EFEFEF}                                                                                                                   & \cellcolor[HTML]{EFEFEF}I                                 & \multicolumn{1}{c|}{\cellcolor[HTML]{EFEFEF}2}          & \multicolumn{1}{c|}{\cellcolor[HTML]{EFEFEF}2}          & \multicolumn{1}{c|}{\cellcolor[HTML]{EFEFEF}1}          & \multicolumn{1}{c|}{\cellcolor[HTML]{EFEFEF}1}          & \multicolumn{1}{c|}{\cellcolor[HTML]{EFEFEF}1}          & \cellcolor[HTML]{CBFE9B}\textbf{7}  \\ \cline{4-10} 
\multirow{-7}{*}{End Devices}                                                                                       & \multirow{-7}{*}{\begin{tabular}[c]{@{}c@{}}ECUs with Sensors \\ and Actuators\end{tabular}}                                         & \multirow{-5}{*}{\cellcolor[HTML]{EFEFEF}\begin{tabular}[c]{@{}c@{}}Software/Firmware change \\ sending compromised messages\end{tabular}} & \cellcolor[HTML]{EFEFEF}D                                 & \multicolumn{1}{c|}{\cellcolor[HTML]{EFEFEF}3}          & \multicolumn{1}{c|}{\cellcolor[HTML]{EFEFEF}1}          & \multicolumn{1}{c|}{\cellcolor[HTML]{EFEFEF}1}          & \multicolumn{1}{c|}{\cellcolor[HTML]{EFEFEF}3}          & \multicolumn{1}{c|}{\cellcolor[HTML]{EFEFEF}1}          & \cellcolor[HTML]{FFFC9E}\textbf{9}  \\ \hline
\cellcolor[HTML]{EFEFEF}                                                                                            & \cellcolor[HTML]{EFEFEF}                                                                                                             &                                                                                                                                            & S                                                         & \multicolumn{1}{c|}{2}                                  & \multicolumn{1}{c|}{3}                                  & \multicolumn{1}{c|}{3}                                  & \multicolumn{1}{c|}{2}                                  & \multicolumn{1}{c|}{2}                                  & \cellcolor[HTML]{FD6864}\textbf{12} \\ \cline{4-10} 
\cellcolor[HTML]{EFEFEF}                                                                                            & \cellcolor[HTML]{EFEFEF}                                                                                                             &                                                                                                                                            & R                                                         & \multicolumn{1}{c|}{2}                                  & \multicolumn{1}{c|}{2}                                  & \multicolumn{1}{c|}{2}                                  & \multicolumn{1}{c|}{2}                                  & \multicolumn{1}{c|}{2}                                  & \cellcolor[HTML]{FFFC9E}\textbf{10} \\ \cline{4-10} 
\cellcolor[HTML]{EFEFEF}                                                                                            & \cellcolor[HTML]{EFEFEF}                                                                                                             &                                                                                                                                            & I                                                         & \multicolumn{1}{c|}{3}                                  & \multicolumn{1}{c|}{2}                                  & \multicolumn{1}{c|}{2}                                  & \multicolumn{1}{c|}{3}                                  & \multicolumn{1}{c|}{2}                                  & \cellcolor[HTML]{FD6864}\textbf{12} \\ \cline{4-10} 
\cellcolor[HTML]{EFEFEF}                                                                                            & \cellcolor[HTML]{EFEFEF}                                                                                                             & \multirow{-4}{*}{Malicious packet spoofing}                                                                                                & D                                                         & \multicolumn{1}{c|}{3}                                  & \multicolumn{1}{c|}{2}                                  & \multicolumn{1}{c|}{2}                                  & \multicolumn{1}{c|}{3}                                  & \multicolumn{1}{c|}{2}                                  & \cellcolor[HTML]{FD6864}\textbf{12} \\ \cline{3-10} 
\rowcolor[HTML]{EFEFEF} 
\cellcolor[HTML]{EFEFEF}                                                                                            & \cellcolor[HTML]{EFEFEF}                                                                                                             & \cellcolor[HTML]{EFEFEF}                                                                                                                   & S                                                         & \multicolumn{1}{c|}{\cellcolor[HTML]{EFEFEF}3}          & \multicolumn{1}{c|}{\cellcolor[HTML]{EFEFEF}3}          & \multicolumn{1}{c|}{\cellcolor[HTML]{EFEFEF}3}          & \multicolumn{1}{c|}{\cellcolor[HTML]{EFEFEF}2}          & \multicolumn{1}{c|}{\cellcolor[HTML]{EFEFEF}3}          & \cellcolor[HTML]{FD6864}\textbf{14} \\ \cline{4-10} 
\rowcolor[HTML]{EFEFEF} 
\cellcolor[HTML]{EFEFEF}                                                                                            & \cellcolor[HTML]{EFEFEF}                                                                                                             & \multirow{-2}{*}{\cellcolor[HTML]{EFEFEF}Denial of Service}                                                                                & D                                                         & \multicolumn{1}{c|}{\cellcolor[HTML]{EFEFEF}3}          & \multicolumn{1}{c|}{\cellcolor[HTML]{EFEFEF}2}          & \multicolumn{1}{c|}{\cellcolor[HTML]{EFEFEF}2}          & \multicolumn{1}{c|}{\cellcolor[HTML]{EFEFEF}3}          & \multicolumn{1}{c|}{\cellcolor[HTML]{EFEFEF}2}          & \cellcolor[HTML]{FD6864}\textbf{12} \\ \cline{3-10} 
\cellcolor[HTML]{EFEFEF}                                                                                            & \cellcolor[HTML]{EFEFEF}                                                                                                             &                                                                                                                                            & S                                                         & \multicolumn{1}{c|}{3}                                  & \multicolumn{1}{c|}{3}                                  & \multicolumn{1}{c|}{3}                                  & \multicolumn{1}{c|}{2}                                  & \multicolumn{1}{c|}{1}                                  & \cellcolor[HTML]{FD6864}\textbf{12} \\ \cline{4-10} 
\cellcolor[HTML]{EFEFEF}                                                                                            & \cellcolor[HTML]{EFEFEF}                                                                                                             &                                                                                                                                            & T                                                         & \multicolumn{1}{c|}{3}                                  & \multicolumn{1}{c|}{3}                                  & \multicolumn{1}{c|}{3}                                  & \multicolumn{1}{c|}{2}                                  & \multicolumn{1}{c|}{1}                                  & \cellcolor[HTML]{FD6864}\textbf{12} \\ \cline{4-10} 
\cellcolor[HTML]{EFEFEF}                                                                                            & \cellcolor[HTML]{EFEFEF}                                                                                                             &                                                                                                                                            & R                                                         & \multicolumn{1}{c|}{2}                                  & \multicolumn{1}{c|}{3}                                  & \multicolumn{1}{c|}{3}                                  & \multicolumn{1}{c|}{2}                                  & \multicolumn{1}{c|}{1}                                  & \cellcolor[HTML]{FFFC9E}\textbf{11} \\ \cline{4-10} 
\cellcolor[HTML]{EFEFEF}                                                                                            & \cellcolor[HTML]{EFEFEF}                                                                                                             & \multirow{-4}{*}{Replay Attacks}                                                                                                           & D                                                         & \multicolumn{1}{c|}{3}                                  & \multicolumn{1}{c|}{2}                                  & \multicolumn{1}{c|}{2}                                  & \multicolumn{1}{c|}{3}                                  & \multicolumn{1}{c|}{1}                                  & \cellcolor[HTML]{FFFC9E}\textbf{11} \\ \cline{3-10} 
\rowcolor[HTML]{EFEFEF} 
\cellcolor[HTML]{EFEFEF}                                                                                            & \cellcolor[HTML]{EFEFEF}                                                                                                             & \cellcolor[HTML]{EFEFEF}                                                                                                                   & S                                                         & \multicolumn{1}{c|}{\cellcolor[HTML]{EFEFEF}3}          & \multicolumn{1}{c|}{\cellcolor[HTML]{EFEFEF}3}          & \multicolumn{1}{c|}{\cellcolor[HTML]{EFEFEF}3}          & \multicolumn{1}{c|}{\cellcolor[HTML]{EFEFEF}2}          & \multicolumn{1}{c|}{\cellcolor[HTML]{EFEFEF}1}          & \cellcolor[HTML]{FD6864}\textbf{12} \\ \cline{4-10} 
\rowcolor[HTML]{EFEFEF} 
\cellcolor[HTML]{EFEFEF}                                                                                            & \cellcolor[HTML]{EFEFEF}                                                                                                             & \cellcolor[HTML]{EFEFEF}                                                                                                                   & R                                                         & \multicolumn{1}{c|}{\cellcolor[HTML]{EFEFEF}3}          & \multicolumn{1}{c|}{\cellcolor[HTML]{EFEFEF}2}          & \multicolumn{1}{c|}{\cellcolor[HTML]{EFEFEF}2}          & \multicolumn{1}{c|}{\cellcolor[HTML]{EFEFEF}2}          & \multicolumn{1}{c|}{\cellcolor[HTML]{EFEFEF}1}          & \cellcolor[HTML]{FFFC9E}\textbf{10} \\ \cline{4-10} 
\rowcolor[HTML]{EFEFEF} 
\cellcolor[HTML]{EFEFEF}                                                                                            & \cellcolor[HTML]{EFEFEF}                                                                                                             & \multirow{-3}{*}{\cellcolor[HTML]{EFEFEF}\begin{tabular}[c]{@{}c@{}}Fuzzing attacks \\ (both ID and data)\end{tabular}}                    & D                                                         & \multicolumn{1}{c|}{\cellcolor[HTML]{EFEFEF}3}          & \multicolumn{1}{c|}{\cellcolor[HTML]{EFEFEF}3}          & \multicolumn{1}{c|}{\cellcolor[HTML]{EFEFEF}3}          & \multicolumn{1}{c|}{\cellcolor[HTML]{EFEFEF}3}          & \multicolumn{1}{c|}{\cellcolor[HTML]{EFEFEF}1}          & \cellcolor[HTML]{FD6864}\textbf{13} \\ \cline{3-10} 
\cellcolor[HTML]{EFEFEF}                                                                                            & \cellcolor[HTML]{EFEFEF}                                                                                                             &                                                                                                                                            & T                                                         & \multicolumn{1}{c|}{3}                                  & \multicolumn{1}{c|}{3}                                  & \multicolumn{1}{c|}{3}                                  & \multicolumn{1}{c|}{2}                                  & \multicolumn{1}{c|}{1}                                  & \cellcolor[HTML]{FD6864}\textbf{12} \\ \cline{4-10} 
\cellcolor[HTML]{EFEFEF}                                                                                            & \cellcolor[HTML]{EFEFEF}                                                                                                             &                                                                                                                                            & R                                                         & \multicolumn{1}{c|}{3}                                  & \multicolumn{1}{c|}{2}                                  & \multicolumn{1}{c|}{3}                                  & \multicolumn{1}{c|}{1}                                  & \multicolumn{1}{c|}{1}                                  & \cellcolor[HTML]{FFFC9E}\textbf{10} \\ \cline{4-10} 
\cellcolor[HTML]{EFEFEF}                                                                                            & \multirow{-16}{*}{\cellcolor[HTML]{EFEFEF}\begin{tabular}[c]{@{}c@{}}Vehicle communication\\ \\ buses like the CAN bus\end{tabular}} & \multirow{-3}{*}{Suspension Attacks}                                                                                                       & D                                                         & \multicolumn{1}{c|}{3}                                  & \multicolumn{1}{c|}{3}                                  & \multicolumn{1}{c|}{3}                                  & \multicolumn{1}{c|}{1}                                  & \multicolumn{1}{c|}{1}                                  & \cellcolor[HTML]{FFFC9E}\textbf{11} \\ \cline{2-10} 
\cellcolor[HTML]{EFEFEF}                                                                                            &                                                                                                                                      & \cellcolor[HTML]{EFEFEF}                                                                                                                   & \cellcolor[HTML]{EFEFEF}T                                 & \multicolumn{1}{c|}{\cellcolor[HTML]{EFEFEF}3}          & \multicolumn{1}{c|}{\cellcolor[HTML]{EFEFEF}3}          & \multicolumn{1}{c|}{\cellcolor[HTML]{EFEFEF}3}          & \multicolumn{1}{c|}{\cellcolor[HTML]{EFEFEF}1}          & \multicolumn{1}{c|}{\cellcolor[HTML]{EFEFEF}3}          & \cellcolor[HTML]{FD6864}\textbf{13} \\ \cline{4-10} 
\cellcolor[HTML]{EFEFEF}                                                                                            &                                                                                                                                      & \multirow{-2}{*}{\cellcolor[HTML]{EFEFEF}Physical Damage}                                                                                  & \cellcolor[HTML]{EFEFEF}D                                 & \multicolumn{1}{c|}{\cellcolor[HTML]{EFEFEF}3}          & \multicolumn{1}{c|}{\cellcolor[HTML]{EFEFEF}3}          & \multicolumn{1}{c|}{\cellcolor[HTML]{EFEFEF}3}          & \multicolumn{1}{c|}{\cellcolor[HTML]{EFEFEF}1}          & \multicolumn{1}{c|}{\cellcolor[HTML]{EFEFEF}3}          & \cellcolor[HTML]{FD6864}\textbf{13} \\ \cline{3-10} 
\cellcolor[HTML]{EFEFEF}                                                                                            &                                                                                                                                      &                                                                                                                                            & S                                                         & \multicolumn{1}{c|}{3}                                  & \multicolumn{1}{c|}{1}                                  & \multicolumn{1}{c|}{1}                                  & \multicolumn{1}{c|}{1}                                  & \multicolumn{1}{c|}{1}                                  & \cellcolor[HTML]{CBFE9B}\textbf{7}  \\ \cline{4-10} 
\cellcolor[HTML]{EFEFEF}                                                                                            &                                                                                                                                      &                                                                                                                                            & T                                                         & \multicolumn{1}{c|}{3}                                  & \multicolumn{1}{c|}{1}                                  & \multicolumn{1}{c|}{1}                                  & \multicolumn{1}{c|}{2}                                  & \multicolumn{1}{c|}{1}                                  & \cellcolor[HTML]{FFFC9E}\textbf{8}  \\ \cline{4-10} 
\cellcolor[HTML]{EFEFEF}                                                                                            &                                                                                                                                      &                                                                                                                                            & R                                                         & \multicolumn{1}{c|}{2}                                  & \multicolumn{1}{c|}{1}                                  & \multicolumn{1}{c|}{1}                                  & \multicolumn{1}{c|}{1}                                  & \multicolumn{1}{c|}{1}                                  & \cellcolor[HTML]{CBFE9B}\textbf{6}  \\ \cline{4-10} 
\cellcolor[HTML]{EFEFEF}                                                                                            &                                                                                                                                      &                                                                                                                                            & D                                                         & \multicolumn{1}{c|}{3}                                  & \multicolumn{1}{c|}{2}                                  & \multicolumn{1}{c|}{2}                                  & \multicolumn{1}{c|}{1}                                  & \multicolumn{1}{c|}{1}                                  & \cellcolor[HTML]{FFFC9E}\textbf{9}  \\ \cline{4-10} 
\cellcolor[HTML]{EFEFEF}                                                                                            &                                                                                                                                      & \multirow{-5}{*}{Firmware changes}                                                                                                         & E                                                         & \multicolumn{1}{c|}{3}                                  & \multicolumn{1}{c|}{1}                                  & \multicolumn{1}{c|}{1}                                  & \multicolumn{1}{c|}{1}                                  & \multicolumn{1}{c|}{1}                                  & \cellcolor[HTML]{CBFE9B}\textbf{7}  \\ \cline{3-10} 
\cellcolor[HTML]{EFEFEF}                                                                                            &                                                                                                                                      & \cellcolor[HTML]{EFEFEF}                                                                                                                   & \cellcolor[HTML]{EFEFEF}S                                 & \multicolumn{1}{c|}{\cellcolor[HTML]{EFEFEF}3}          & \multicolumn{1}{c|}{\cellcolor[HTML]{EFEFEF}1}          & \multicolumn{1}{c|}{\cellcolor[HTML]{EFEFEF}1}          & \multicolumn{1}{c|}{\cellcolor[HTML]{EFEFEF}1}          & \multicolumn{1}{c|}{\cellcolor[HTML]{EFEFEF}1}          & \cellcolor[HTML]{CBFE9B}\textbf{7}  \\ \cline{4-10} 
\cellcolor[HTML]{EFEFEF}                                                                                            &                                                                                                                                      & \cellcolor[HTML]{EFEFEF}                                                                                                                   & \cellcolor[HTML]{EFEFEF}T                                 & \multicolumn{1}{c|}{\cellcolor[HTML]{EFEFEF}3}          & \multicolumn{1}{c|}{\cellcolor[HTML]{EFEFEF}1}          & \multicolumn{1}{c|}{\cellcolor[HTML]{EFEFEF}1}          & \multicolumn{1}{c|}{\cellcolor[HTML]{EFEFEF}1}          & \multicolumn{1}{c|}{\cellcolor[HTML]{EFEFEF}1}          & \cellcolor[HTML]{CBFE9B}\textbf{7}  \\ \cline{4-10} 
\cellcolor[HTML]{EFEFEF}                                                                                            &                                                                                                                                      & \cellcolor[HTML]{EFEFEF}                                                                                                                   & \cellcolor[HTML]{EFEFEF}R                                 & \multicolumn{1}{c|}{\cellcolor[HTML]{EFEFEF}3}          & \multicolumn{1}{c|}{\cellcolor[HTML]{EFEFEF}1}          & \multicolumn{1}{c|}{\cellcolor[HTML]{EFEFEF}1}          & \multicolumn{1}{c|}{\cellcolor[HTML]{EFEFEF}1}          & \multicolumn{1}{c|}{\cellcolor[HTML]{EFEFEF}1}          & \cellcolor[HTML]{CBFE9B}\textbf{7}  \\ \cline{4-10} 
\cellcolor[HTML]{EFEFEF}                                                                                            &                                                                                                                                      & \cellcolor[HTML]{EFEFEF}                                                                                                                   & \cellcolor[HTML]{EFEFEF}D                                 & \multicolumn{1}{c|}{\cellcolor[HTML]{EFEFEF}3}          & \multicolumn{1}{c|}{\cellcolor[HTML]{EFEFEF}2}          & \multicolumn{1}{c|}{\cellcolor[HTML]{EFEFEF}1}          & \multicolumn{1}{c|}{\cellcolor[HTML]{EFEFEF}1}          & \multicolumn{1}{c|}{\cellcolor[HTML]{EFEFEF}1}          & \cellcolor[HTML]{FFFC9E}\textbf{8}  \\ \cline{4-10} 
\cellcolor[HTML]{EFEFEF}                                                                                            & \multirow{-12}{*}{Head Unit}                                                                                                         & \multirow{-5}{*}{\cellcolor[HTML]{EFEFEF}Installation of malicious apps}                                                                   & \cellcolor[HTML]{EFEFEF}E                                 & \multicolumn{1}{c|}{\cellcolor[HTML]{EFEFEF}3}          & \multicolumn{1}{c|}{\cellcolor[HTML]{EFEFEF}1}          & \multicolumn{1}{c|}{\cellcolor[HTML]{EFEFEF}1}          & \multicolumn{1}{c|}{\cellcolor[HTML]{EFEFEF}1}          & \multicolumn{1}{c|}{\cellcolor[HTML]{EFEFEF}1}          & \cellcolor[HTML]{CBFE9B}\textbf{7}  \\ \cline{2-10} 
\cellcolor[HTML]{EFEFEF}                                                                                            & \cellcolor[HTML]{EFEFEF}                                                                                                             &                                                                                                                                            & T                                                         & \multicolumn{1}{c|}{3}                                  & \multicolumn{1}{c|}{3}                                  & \multicolumn{1}{c|}{3}                                  & \multicolumn{1}{c|}{1}                                  & \multicolumn{1}{c|}{3}                                  & \cellcolor[HTML]{FD6864}\textbf{13} \\ \cline{4-10} 
\cellcolor[HTML]{EFEFEF}                                                                                            & \cellcolor[HTML]{EFEFEF}                                                                                                             & \multirow{-2}{*}{Physical Damage}                                                                                                          & D                                                         & \multicolumn{1}{c|}{3}                                  & \multicolumn{1}{c|}{3}                                  & \multicolumn{1}{c|}{3}                                  & \multicolumn{1}{c|}{2}                                  & \multicolumn{1}{c|}{3}                                  & \cellcolor[HTML]{FD6864}\textbf{14} \\ \cline{3-10} 
\rowcolor[HTML]{EFEFEF} 
\cellcolor[HTML]{EFEFEF}                                                                                            & \cellcolor[HTML]{EFEFEF}                                                                                                             & \cellcolor[HTML]{EFEFEF}                                                                                                                   & S                                                         & \multicolumn{1}{c|}{\cellcolor[HTML]{EFEFEF}3}          & \multicolumn{1}{c|}{\cellcolor[HTML]{EFEFEF}1}          & \multicolumn{1}{c|}{\cellcolor[HTML]{EFEFEF}1}          & \multicolumn{1}{c|}{\cellcolor[HTML]{EFEFEF}1}          & \multicolumn{1}{c|}{\cellcolor[HTML]{EFEFEF}1}          & \cellcolor[HTML]{CBFE9B}\textbf{7}  \\ \cline{4-10} 
\rowcolor[HTML]{EFEFEF} 
\cellcolor[HTML]{EFEFEF}                                                                                            & \cellcolor[HTML]{EFEFEF}                                                                                                             & \cellcolor[HTML]{EFEFEF}                                                                                                                   & T                                                         & \multicolumn{1}{c|}{\cellcolor[HTML]{EFEFEF}3}          & \multicolumn{1}{c|}{\cellcolor[HTML]{EFEFEF}1}          & \multicolumn{1}{c|}{\cellcolor[HTML]{EFEFEF}1}          & \multicolumn{1}{c|}{\cellcolor[HTML]{EFEFEF}1}          & \multicolumn{1}{c|}{\cellcolor[HTML]{EFEFEF}1}          & \cellcolor[HTML]{CBFE9B}\textbf{7}  \\ \cline{4-10} 
\rowcolor[HTML]{EFEFEF} 
\cellcolor[HTML]{EFEFEF}                                                                                            & \cellcolor[HTML]{EFEFEF}                                                                                                             & \cellcolor[HTML]{EFEFEF}                                                                                                                   & R                                                         & \multicolumn{1}{c|}{\cellcolor[HTML]{EFEFEF}3}          & \multicolumn{1}{c|}{\cellcolor[HTML]{EFEFEF}1}          & \multicolumn{1}{c|}{\cellcolor[HTML]{EFEFEF}1}          & \multicolumn{1}{c|}{\cellcolor[HTML]{EFEFEF}1}          & \multicolumn{1}{c|}{\cellcolor[HTML]{EFEFEF}1}          & \cellcolor[HTML]{CBFE9B}\textbf{7}  \\ \cline{4-10} 
\rowcolor[HTML]{EFEFEF} 
\cellcolor[HTML]{EFEFEF}                                                                                            & \cellcolor[HTML]{EFEFEF}                                                                                                             & \cellcolor[HTML]{EFEFEF}                                                                                                                   & D                                                         & \multicolumn{1}{c|}{\cellcolor[HTML]{EFEFEF}3}          & \multicolumn{1}{c|}{\cellcolor[HTML]{EFEFEF}2}          & \multicolumn{1}{c|}{\cellcolor[HTML]{EFEFEF}2}          & \multicolumn{1}{c|}{\cellcolor[HTML]{EFEFEF}2}          & \multicolumn{1}{c|}{\cellcolor[HTML]{EFEFEF}1}          & \cellcolor[HTML]{FFFC9E}\textbf{10} \\ \cline{4-10} 
\rowcolor[HTML]{EFEFEF} 
\multirow{-35}{*}{\cellcolor[HTML]{EFEFEF}\begin{tabular}[c]{@{}c@{}}Intra Vehicle \\ Networks\end{tabular}}        & \multirow{-7}{*}{\cellcolor[HTML]{EFEFEF}\begin{tabular}[c]{@{}c@{}}Ethernet Gateway \\ and OBDII\end{tabular}}                      & \multirow{-5}{*}{\cellcolor[HTML]{EFEFEF}Firmware change}                                                                                  & E                                                         & \multicolumn{1}{c|}{\cellcolor[HTML]{EFEFEF}3}          & \multicolumn{1}{c|}{\cellcolor[HTML]{EFEFEF}2}          & \multicolumn{1}{c|}{\cellcolor[HTML]{EFEFEF}1}          & \multicolumn{1}{c|}{\cellcolor[HTML]{EFEFEF}1}          & \multicolumn{1}{c|}{\cellcolor[HTML]{EFEFEF}1}          & \cellcolor[HTML]{FFFC9E}\textbf{8}  \\ \hline
                                                                                                                    &                                                                                                                                      &                                                                                                                                            & S                                                         & \multicolumn{1}{c|}{2}                                  & \multicolumn{1}{c|}{3}                                  & \multicolumn{1}{c|}{2}                                  & \multicolumn{1}{c|}{2}                                  & \multicolumn{1}{c|}{3}                                  & \cellcolor[HTML]{FD6864}\textbf{12} \\ \cline{4-10} 
                                                                                                                    &                                                                                                                                      &                                                                                                                                            & R                                                         & \multicolumn{1}{c|}{2}                                  & \multicolumn{1}{c|}{3}                                  & \multicolumn{1}{c|}{2}                                  & \multicolumn{1}{c|}{2}                                  & \multicolumn{1}{c|}{3}                                  & \cellcolor[HTML]{FD6864}\textbf{12} \\ \cline{4-10} 
                                                                                                                    &                                                                                                                                      &                                                                                                                                            & I                                                         & \multicolumn{1}{c|}{2}                                  & \multicolumn{1}{c|}{3}                                  & \multicolumn{1}{c|}{2}                                  & \multicolumn{1}{c|}{2}                                  & \multicolumn{1}{c|}{3}                                  & \cellcolor[HTML]{FD6864}\textbf{12} \\ \cline{4-10} 
                                                                                                                    &                                                                                                                                      & \multirow{-4}{*}{Frame Spoofing}                                                                                                           & D                                                         & \multicolumn{1}{c|}{3}                                  & \multicolumn{1}{c|}{2}                                  & \multicolumn{1}{c|}{3}                                  & \multicolumn{1}{c|}{2}                                  & \multicolumn{1}{c|}{3}                                  & \cellcolor[HTML]{FD6864}\textbf{13} \\ \cline{3-10} 
                                                                                                                    &                                                                                                                                      & \cellcolor[HTML]{EFEFEF}                                                                                                                   & \cellcolor[HTML]{EFEFEF}S                                 & \multicolumn{1}{c|}{\cellcolor[HTML]{EFEFEF}2}          & \multicolumn{1}{c|}{\cellcolor[HTML]{EFEFEF}3}          & \multicolumn{1}{c|}{\cellcolor[HTML]{EFEFEF}2}          & \multicolumn{1}{c|}{\cellcolor[HTML]{EFEFEF}2}          & \multicolumn{1}{c|}{\cellcolor[HTML]{EFEFEF}3}          & \cellcolor[HTML]{FD6864}\textbf{12} \\ \cline{4-10} 
                                                                                                                    &                                                                                                                                      & \multirow{-2}{*}{\cellcolor[HTML]{EFEFEF}Unauthorized Deauthentication}                                                                    & \cellcolor[HTML]{EFEFEF}D                                 & \multicolumn{1}{c|}{\cellcolor[HTML]{EFEFEF}3}          & \multicolumn{1}{c|}{\cellcolor[HTML]{EFEFEF}3}          & \multicolumn{1}{c|}{\cellcolor[HTML]{EFEFEF}2}          & \multicolumn{1}{c|}{\cellcolor[HTML]{EFEFEF}3}          & \multicolumn{1}{c|}{\cellcolor[HTML]{EFEFEF}2}          & \cellcolor[HTML]{FD6864}\textbf{13} \\ \cline{3-10} 
                                                                                                                    &                                                                                                                                      &                                                                                                                                            & S                                                         & \multicolumn{1}{c|}{3}                                  & \multicolumn{1}{c|}{3}                                  & \multicolumn{1}{c|}{3}                                  & \multicolumn{1}{c|}{2}                                  & \multicolumn{1}{c|}{3}                                  & \cellcolor[HTML]{FD6864}\textbf{14} \\ \cline{4-10} 
                                                                                                                    &                                                                                                                                      &                                                                                                                                            & T                                                         & \multicolumn{1}{c|}{3}                                  & \multicolumn{1}{c|}{2}                                  & \multicolumn{1}{c|}{3}                                  & \multicolumn{1}{c|}{2}                                  & \multicolumn{1}{c|}{2}                                  & \cellcolor[HTML]{FD6864}\textbf{12} \\ \cline{4-10} 
                                                                                                                    &                                                                                                                                      &                                                                                                                                            & R                                                         & \multicolumn{1}{c|}{2}                                  & \multicolumn{1}{c|}{3}                                  & \multicolumn{1}{c|}{3}                                  & \multicolumn{1}{c|}{2}                                  & \multicolumn{1}{c|}{3}                                  & \cellcolor[HTML]{FD6864}\textbf{13} \\ \cline{4-10} 
                                                                                                                    &                                                                                                                                      & \multirow{-4}{*}{Fake Access Points}                                                                                                       & D                                                         & \multicolumn{1}{c|}{3}                                  & \multicolumn{1}{c|}{2}                                  & \multicolumn{1}{c|}{2}                                  & \multicolumn{1}{c|}{2}                                  & \multicolumn{1}{c|}{3}                                  & \cellcolor[HTML]{FD6864}\textbf{12} \\ \cline{3-10} 
                                                                                                                    &                                                                                                                                      & \cellcolor[HTML]{EFEFEF}                                                                                                                   & \cellcolor[HTML]{EFEFEF}S                                 & \multicolumn{1}{c|}{\cellcolor[HTML]{EFEFEF}2}          & \multicolumn{1}{c|}{\cellcolor[HTML]{EFEFEF}3}          & \multicolumn{1}{c|}{\cellcolor[HTML]{EFEFEF}2}          & \multicolumn{1}{c|}{\cellcolor[HTML]{EFEFEF}2}          & \multicolumn{1}{c|}{\cellcolor[HTML]{EFEFEF}3}          & \cellcolor[HTML]{FD6864}\textbf{12} \\ \cline{4-10} 
                                                                                                                    & \multirow{-12}{*}{WiFi}                                                                                                              & \multirow{-2}{*}{\cellcolor[HTML]{EFEFEF}Denial of Service}                                                                                & \cellcolor[HTML]{EFEFEF}D                                 & \multicolumn{1}{c|}{\cellcolor[HTML]{EFEFEF}3}          & \multicolumn{1}{c|}{\cellcolor[HTML]{EFEFEF}2}          & \multicolumn{1}{c|}{\cellcolor[HTML]{EFEFEF}3}          & \multicolumn{1}{c|}{\cellcolor[HTML]{EFEFEF}3}          & \multicolumn{1}{c|}{\cellcolor[HTML]{EFEFEF}2}          & \cellcolor[HTML]{FD6864}\textbf{13} \\ \cline{2-10} 
                                                                                                                    & \cellcolor[HTML]{EFEFEF}                                                                                                             &                                                                                                                                            & I                                                         & \multicolumn{1}{c|}{2}                                  & \multicolumn{1}{c|}{1}                                  & \multicolumn{1}{c|}{3}                                  & \multicolumn{1}{c|}{3}                                  & \multicolumn{1}{c|}{3}                                  & \cellcolor[HTML]{FD6864}\textbf{12} \\ \cline{4-10} 
                                                                                                                    & \cellcolor[HTML]{EFEFEF}                                                                                                             & \multirow{-2}{*}{Packet Sniffing}                                                                                                          & E                                                         & \multicolumn{1}{c|}{1}                                  & \multicolumn{1}{c|}{1}                                  & \multicolumn{1}{c|}{3}                                  & \multicolumn{1}{c|}{3}                                  & \multicolumn{1}{c|}{3}                                  & \cellcolor[HTML]{FFFC9E}\textbf{11} \\ \cline{3-10} 
                                                                                                                    & \cellcolor[HTML]{EFEFEF}                                                                                                             & \cellcolor[HTML]{EFEFEF}                                                                                                                   & \cellcolor[HTML]{EFEFEF}S                                 & \multicolumn{1}{c|}{\cellcolor[HTML]{EFEFEF}2}          & \multicolumn{1}{c|}{\cellcolor[HTML]{EFEFEF}3}          & \multicolumn{1}{c|}{\cellcolor[HTML]{EFEFEF}3}          & \multicolumn{1}{c|}{\cellcolor[HTML]{EFEFEF}3}          & \multicolumn{1}{c|}{\cellcolor[HTML]{EFEFEF}3}          & \cellcolor[HTML]{FD6864}\textbf{14} \\ \cline{4-10} 
                                                                                                                    & \cellcolor[HTML]{EFEFEF}                                                                                                             & \cellcolor[HTML]{EFEFEF}                                                                                                                   & \cellcolor[HTML]{EFEFEF}T                                 & \multicolumn{1}{c|}{\cellcolor[HTML]{EFEFEF}3}          & \multicolumn{1}{c|}{\cellcolor[HTML]{EFEFEF}3}          & \multicolumn{1}{c|}{\cellcolor[HTML]{EFEFEF}2}          & \multicolumn{1}{c|}{\cellcolor[HTML]{EFEFEF}3}          & \multicolumn{1}{c|}{\cellcolor[HTML]{EFEFEF}3}          & \cellcolor[HTML]{FD6864}\textbf{14} \\ \cline{4-10} 
                                                                                                                    & \cellcolor[HTML]{EFEFEF}                                                                                                             & \cellcolor[HTML]{EFEFEF}                                                                                                                   & \cellcolor[HTML]{EFEFEF}R                                 & \multicolumn{1}{c|}{\cellcolor[HTML]{EFEFEF}3}          & \multicolumn{1}{c|}{\cellcolor[HTML]{EFEFEF}3}          & \multicolumn{1}{c|}{\cellcolor[HTML]{EFEFEF}2}          & \multicolumn{1}{c|}{\cellcolor[HTML]{EFEFEF}3}          & \multicolumn{1}{c|}{\cellcolor[HTML]{EFEFEF}3}          & \cellcolor[HTML]{FD6864}\textbf{14} \\ \cline{4-10} 
                                                                                                                    & \cellcolor[HTML]{EFEFEF}                                                                                                             & \cellcolor[HTML]{EFEFEF}                                                                                                                   & \cellcolor[HTML]{EFEFEF}I                                 & \multicolumn{1}{c|}{\cellcolor[HTML]{EFEFEF}2}          & \multicolumn{1}{c|}{\cellcolor[HTML]{EFEFEF}3}          & \multicolumn{1}{c|}{\cellcolor[HTML]{EFEFEF}2}          & \multicolumn{1}{c|}{\cellcolor[HTML]{EFEFEF}3}          & \multicolumn{1}{c|}{\cellcolor[HTML]{EFEFEF}3}          & \cellcolor[HTML]{FD6864}\textbf{13} \\ \cline{4-10} 
                                                                                                                    & \cellcolor[HTML]{EFEFEF}                                                                                                             & \cellcolor[HTML]{EFEFEF}                                                                                                                   & \cellcolor[HTML]{EFEFEF}D                                 & \multicolumn{1}{c|}{\cellcolor[HTML]{EFEFEF}3}          & \multicolumn{1}{c|}{\cellcolor[HTML]{EFEFEF}3}          & \multicolumn{1}{c|}{\cellcolor[HTML]{EFEFEF}3}          & \multicolumn{1}{c|}{\cellcolor[HTML]{EFEFEF}3}          & \multicolumn{1}{c|}{\cellcolor[HTML]{EFEFEF}3}          & \cellcolor[HTML]{FD6864}\textbf{15} \\ \cline{4-10} 
                                                                                                                    & \cellcolor[HTML]{EFEFEF}                                                                                                             & \multirow{-6}{*}{\cellcolor[HTML]{EFEFEF}Man in the Middle}                                                                                & \cellcolor[HTML]{EFEFEF}E                                 & \multicolumn{1}{c|}{\cellcolor[HTML]{EFEFEF}2}          & \multicolumn{1}{c|}{\cellcolor[HTML]{EFEFEF}3}          & \multicolumn{1}{c|}{\cellcolor[HTML]{EFEFEF}3}          & \multicolumn{1}{c|}{\cellcolor[HTML]{EFEFEF}3}          & \multicolumn{1}{c|}{\cellcolor[HTML]{EFEFEF}3}          & \cellcolor[HTML]{FD6864}\textbf{14} \\ \cline{3-10} 
                                                                                                                    & \multirow{-9}{*}{\cellcolor[HTML]{EFEFEF}Bluetooth}                                                                                  & Battery Draining                                                                                                                           & D                                                         & \multicolumn{1}{c|}{2}                                  & \multicolumn{1}{c|}{3}                                  & \multicolumn{1}{c|}{3}                                  & \multicolumn{1}{c|}{3}                                  & \multicolumn{1}{c|}{3}                                  & \cellcolor[HTML]{FD6864}\textbf{14} \\ \cline{2-10} 
                                                                                                                    &                                                                                                                                      & \cellcolor[HTML]{EFEFEF}                                                                                                                   & \cellcolor[HTML]{EFEFEF}I                                 & \multicolumn{1}{c|}{\cellcolor[HTML]{EFEFEF}2}          & \multicolumn{1}{c|}{\cellcolor[HTML]{EFEFEF}1}          & \multicolumn{1}{c|}{\cellcolor[HTML]{EFEFEF}1}          & \multicolumn{1}{c|}{\cellcolor[HTML]{EFEFEF}3}          & \multicolumn{1}{c|}{\cellcolor[HTML]{EFEFEF}1}          & \cellcolor[HTML]{FFFC9E}\textbf{8}  \\ \cline{4-10} 
                                                                                                                    &                                                                                                                                      & \multirow{-2}{*}{\cellcolor[HTML]{EFEFEF}Packet sniffing}                                                                                  & \cellcolor[HTML]{EFEFEF}E                                 & \multicolumn{1}{c|}{\cellcolor[HTML]{EFEFEF}2}          & \multicolumn{1}{c|}{\cellcolor[HTML]{EFEFEF}2}          & \multicolumn{1}{c|}{\cellcolor[HTML]{EFEFEF}1}          & \multicolumn{1}{c|}{\cellcolor[HTML]{EFEFEF}2}          & \multicolumn{1}{c|}{\cellcolor[HTML]{EFEFEF}1}          & \cellcolor[HTML]{FFFC9E}\textbf{8}  \\ \cline{3-10} 
                                                                                                                    &                                                                                                                                      &                                                                                                                                            & \cellcolor[HTML]{FFFFFF}S                                 & \multicolumn{1}{c|}{3}                                  & \multicolumn{1}{c|}{2}                                  & \multicolumn{1}{c|}{2}                                  & \multicolumn{1}{c|}{3}                                  & \multicolumn{1}{c|}{1}                                  & \cellcolor[HTML]{FFFC9E}\textbf{11} \\ \cline{4-10} 
                                                                                                                    &                                                                                                                                      &                                                                                                                                            & T                                                         & \multicolumn{1}{c|}{2}                                  & \multicolumn{1}{c|}{2}                                  & \multicolumn{1}{c|}{3}                                  & \multicolumn{1}{c|}{3}                                  & \multicolumn{1}{c|}{1}                                  & \cellcolor[HTML]{FFFC9E}\textbf{11} \\ \cline{4-10} 
                                                                                                                    &                                                                                                                                      &                                                                                                                                            & R                                                         & \multicolumn{1}{c|}{2}                                  & \multicolumn{1}{c|}{2}                                  & \multicolumn{1}{c|}{3}                                  & \multicolumn{1}{c|}{2}                                  & \multicolumn{1}{c|}{1}                                  & \cellcolor[HTML]{FFFC9E}\textbf{10} \\ \cline{4-10} 
                                                                                                                    &                                                                                                                                      &                                                                                                                                            & I                                                         & \multicolumn{1}{c|}{3}                                  & \multicolumn{1}{c|}{2}                                  & \multicolumn{1}{c|}{2}                                  & \multicolumn{1}{c|}{3}                                  & \multicolumn{1}{c|}{1}                                  & \cellcolor[HTML]{FFFC9E}\textbf{11} \\ \cline{4-10} 
                                                                                                                    &                                                                                                                                      &                                                                                                                                            & D                                                         & \multicolumn{1}{c|}{3}                                  & \multicolumn{1}{c|}{3}                                  & \multicolumn{1}{c|}{3}                                  & \multicolumn{1}{c|}{3}                                  & \multicolumn{1}{c|}{1}                                  & \cellcolor[HTML]{FD6864}\textbf{13} \\ \cline{4-10} 
                                                                                                                    &                                                                                                                                      & \multirow{-6}{*}{Man in the middle}                                                                                                        & E                                                         & \multicolumn{1}{c|}{3}                                  & \multicolumn{1}{c|}{2}                                  & \multicolumn{1}{c|}{3}                                  & \multicolumn{1}{c|}{2}                                  & \multicolumn{1}{c|}{1}                                  & \cellcolor[HTML]{FFFC9E}\textbf{11} \\ \cline{3-10} 
                                                                                                                    &                                                                                                                                      & \cellcolor[HTML]{EFEFEF}                                                                                                                   & \cellcolor[HTML]{EFEFEF}R                                 & \multicolumn{1}{c|}{\cellcolor[HTML]{EFEFEF}3}          & \multicolumn{1}{c|}{\cellcolor[HTML]{EFEFEF}3}          & \multicolumn{1}{c|}{\cellcolor[HTML]{EFEFEF}1}          & \multicolumn{1}{c|}{\cellcolor[HTML]{EFEFEF}3}          & \multicolumn{1}{c|}{\cellcolor[HTML]{EFEFEF}1}          & \cellcolor[HTML]{FFFC9E}\textbf{11} \\ \cline{4-10} 
                                                                                                                    &                                                                                                                                      & \cellcolor[HTML]{EFEFEF}                                                                                                                   & \cellcolor[HTML]{EFEFEF}I                                 & \multicolumn{1}{c|}{\cellcolor[HTML]{EFEFEF}3}          & \multicolumn{1}{c|}{\cellcolor[HTML]{EFEFEF}3}          & \multicolumn{1}{c|}{\cellcolor[HTML]{EFEFEF}2}          & \multicolumn{1}{c|}{\cellcolor[HTML]{EFEFEF}2}          & \multicolumn{1}{c|}{\cellcolor[HTML]{EFEFEF}2}          & \cellcolor[HTML]{FD6864}\textbf{12} \\ \cline{4-10} 
\multirow{-32}{*}{\begin{tabular}[c]{@{}c@{}}Inter Vehicle \\ Networks\end{tabular}}                                & \multirow{-11}{*}{eSIM}                                                                                                              & \multirow{-3}{*}{\cellcolor[HTML]{EFEFEF}Abusing misconfigurations}                                                                        & \cellcolor[HTML]{EFEFEF}E                                 & \multicolumn{1}{c|}{\cellcolor[HTML]{EFEFEF}3}          & \multicolumn{1}{c|}{\cellcolor[HTML]{EFEFEF}2}          & \multicolumn{1}{c|}{\cellcolor[HTML]{EFEFEF}1}          & \multicolumn{1}{c|}{\cellcolor[HTML]{EFEFEF}3}          & \multicolumn{1}{c|}{\cellcolor[HTML]{EFEFEF}1}          & \cellcolor[HTML]{FFFC9E}\textbf{10} \\ \hline
\end{tabular}
\end{table*}

\twocolumn

\begin{table*}[ht!]
\centering
\begin{tabular}{|c|c|c|c|ccccc
>{\columncolor[HTML]{FFFC9E}}c |}
\hline
\cellcolor[HTML]{C0C0C0}                                                                                            & \cellcolor[HTML]{C0C0C0}                                     & \cellcolor[HTML]{C0C0C0}                                            & \cellcolor[HTML]{C0C0C0}                                  & \multicolumn{6}{c|}{\cellcolor[HTML]{C0C0C0}\textbf{Threat Impact}}                                                                                                                                                                                                                                                                       \\ \cline{5-10} 
\multirow{-2}{*}{\cellcolor[HTML]{C0C0C0}\textbf{\begin{tabular}[c]{@{}c@{}}AV Architecture \\ Layer\end{tabular}}} & \multirow{-2}{*}{\cellcolor[HTML]{C0C0C0}\textbf{Asset}}     & \multirow{-2}{*}{\cellcolor[HTML]{C0C0C0}\textbf{Attack Surface}}   & \multirow{-2}{*}{\cellcolor[HTML]{C0C0C0}\textbf{STRIDE}} & \multicolumn{1}{c|}{\cellcolor[HTML]{C0C0C0}\textbf{D}} & \multicolumn{1}{c|}{\cellcolor[HTML]{C0C0C0}\textbf{R}} & \multicolumn{1}{c|}{\cellcolor[HTML]{C0C0C0}\textbf{E}} & \multicolumn{1}{c|}{\cellcolor[HTML]{C0C0C0}\textbf{A}} & \multicolumn{1}{c|}{\cellcolor[HTML]{C0C0C0}\textbf{D}} & \cellcolor[HTML]{C0C0C0}\textbf{Rating} \\ \hline
                                                                                                                    &                                                              & \cellcolor[HTML]{EFEFEF}                                            & \cellcolor[HTML]{EFEFEF}I                                 & \multicolumn{1}{c|}{\cellcolor[HTML]{EFEFEF}2}          & \multicolumn{1}{c|}{\cellcolor[HTML]{EFEFEF}1}          & \multicolumn{1}{c|}{\cellcolor[HTML]{EFEFEF}1}          & \multicolumn{1}{c|}{\cellcolor[HTML]{EFEFEF}3}          & \multicolumn{1}{c|}{\cellcolor[HTML]{EFEFEF}1}          & \textbf{8}                              \\ \cline{4-10} 
                                                                                                                    &                                                              & \multirow{-2}{*}{\cellcolor[HTML]{EFEFEF}Packet sniffing}           & \cellcolor[HTML]{EFEFEF}E                                 & \multicolumn{1}{c|}{\cellcolor[HTML]{EFEFEF}2}          & \multicolumn{1}{c|}{\cellcolor[HTML]{EFEFEF}2}          & \multicolumn{1}{c|}{\cellcolor[HTML]{EFEFEF}1}          & \multicolumn{1}{c|}{\cellcolor[HTML]{EFEFEF}2}          & \multicolumn{1}{c|}{\cellcolor[HTML]{EFEFEF}1}          & \textbf{8}                              \\ \cline{3-10} 
                                                                                                                    &                                                              &                                                                     & \cellcolor[HTML]{FFFFFF}S                                 & \multicolumn{1}{c|}{3}                                  & \multicolumn{1}{c|}{2}                                  & \multicolumn{1}{c|}{2}                                  & \multicolumn{1}{c|}{3}                                  & \multicolumn{1}{c|}{1}                                  & \textbf{11}                             \\ \cline{4-10} 
                                                                                                                    &                                                              &                                                                     & T                                                         & \multicolumn{1}{c|}{2}                                  & \multicolumn{1}{c|}{2}                                  & \multicolumn{1}{c|}{3}                                  & \multicolumn{1}{c|}{3}                                  & \multicolumn{1}{c|}{1}                                  & \textbf{11}                             \\ \cline{4-10} 
                                                                                                                    &                                                              &                                                                     & R                                                         & \multicolumn{1}{c|}{2}                                  & \multicolumn{1}{c|}{2}                                  & \multicolumn{1}{c|}{3}                                  & \multicolumn{1}{c|}{2}                                  & \multicolumn{1}{c|}{1}                                  & \textbf{10}                             \\ \cline{4-10} 
                                                                                                                    &                                                              &                                                                     & I                                                         & \multicolumn{1}{c|}{3}                                  & \multicolumn{1}{c|}{2}                                  & \multicolumn{1}{c|}{2}                                  & \multicolumn{1}{c|}{3}                                  & \multicolumn{1}{c|}{1}                                  & \textbf{11}                             \\ \cline{4-10} 
                                                                                                                    &                                                              &                                                                     & D                                                         & \multicolumn{1}{c|}{3}                                  & \multicolumn{1}{c|}{3}                                  & \multicolumn{1}{c|}{3}                                  & \multicolumn{1}{c|}{3}                                  & \multicolumn{1}{c|}{1}                                  & \cellcolor[HTML]{FD6864}\textbf{13}     \\ \cline{4-10} 
                                                                                                                    &                                                              & \multirow{-6}{*}{Man in the middle}                                 & E                                                         & \multicolumn{1}{c|}{3}                                  & \multicolumn{1}{c|}{2}                                  & \multicolumn{1}{c|}{3}                                  & \multicolumn{1}{c|}{2}                                  & \multicolumn{1}{c|}{1}                                  & \textbf{11}                             \\ \cline{3-10} 
                                                                                                                    &                                                              & \cellcolor[HTML]{EFEFEF}                                            & \cellcolor[HTML]{EFEFEF}R                                 & \multicolumn{1}{c|}{\cellcolor[HTML]{EFEFEF}3}          & \multicolumn{1}{c|}{\cellcolor[HTML]{EFEFEF}3}          & \multicolumn{1}{c|}{\cellcolor[HTML]{EFEFEF}1}          & \multicolumn{1}{c|}{\cellcolor[HTML]{EFEFEF}3}          & \multicolumn{1}{c|}{\cellcolor[HTML]{EFEFEF}1}          & \textbf{11}                             \\ \cline{4-10} 
                                                                                                                    &                                                              & \cellcolor[HTML]{EFEFEF}                                            & \cellcolor[HTML]{EFEFEF}I                                 & \multicolumn{1}{c|}{\cellcolor[HTML]{EFEFEF}3}          & \multicolumn{1}{c|}{\cellcolor[HTML]{EFEFEF}3}          & \multicolumn{1}{c|}{\cellcolor[HTML]{EFEFEF}2}          & \multicolumn{1}{c|}{\cellcolor[HTML]{EFEFEF}2}          & \multicolumn{1}{c|}{\cellcolor[HTML]{EFEFEF}2}          & \cellcolor[HTML]{FD6864}\textbf{12}     \\ \cline{4-10} 
                                                                                                                    & \multirow{-11}{*}{eSIM}                                      & \multirow{-3}{*}{\cellcolor[HTML]{EFEFEF}Abusing misconfigurations} & \cellcolor[HTML]{EFEFEF}E                                 & \multicolumn{1}{c|}{\cellcolor[HTML]{EFEFEF}3}          & \multicolumn{1}{c|}{\cellcolor[HTML]{EFEFEF}2}          & \multicolumn{1}{c|}{\cellcolor[HTML]{EFEFEF}1}          & \multicolumn{1}{c|}{\cellcolor[HTML]{EFEFEF}3}          & \multicolumn{1}{c|}{\cellcolor[HTML]{EFEFEF}1}          & \textbf{10}                             \\ \cline{2-10} 
                                                                                                                    &                                                              &                                                                     & S                                                         & \multicolumn{1}{c|}{3}                                  & \multicolumn{1}{c|}{3}                                  & \multicolumn{1}{c|}{3}                                  & \multicolumn{1}{c|}{3}                                  & \multicolumn{1}{c|}{1}                                  & \cellcolor[HTML]{FD6864}\textbf{13}     \\ \cline{4-10} 
                                                                                                                    &                                                              &                                                                     & R                                                         & \multicolumn{1}{c|}{3}                                  & \multicolumn{1}{c|}{2}                                  & \multicolumn{1}{c|}{2}                                  & \multicolumn{1}{c|}{3}                                  & \multicolumn{1}{c|}{1}                                  & \textbf{11}                             \\ \cline{4-10} 
                                                                                                                    &                                                              &                                                                     & I                                                         & \multicolumn{1}{c|}{2}                                  & \multicolumn{1}{c|}{2}                                  & \multicolumn{1}{c|}{2}                                  & \multicolumn{1}{c|}{2}                                  & \multicolumn{1}{c|}{1}                                  & \textbf{9}                              \\ \cline{4-10} 
                                                                                                                    &                                                              & \multirow{-4}{*}{Spoofing}                                          & D                                                         & \multicolumn{1}{c|}{3}                                  & \multicolumn{1}{c|}{2}                                  & \multicolumn{1}{c|}{2}                                  & \multicolumn{1}{c|}{3}                                  & \multicolumn{1}{c|}{1}                                  & \textbf{11}                             \\ \cline{3-10} 
                                                                                                                    &                                                              & \cellcolor[HTML]{EFEFEF}                                            & \cellcolor[HTML]{EFEFEF}S                                 & \multicolumn{1}{c|}{\cellcolor[HTML]{EFEFEF}3}          & \multicolumn{1}{c|}{\cellcolor[HTML]{EFEFEF}2}          & \multicolumn{1}{c|}{\cellcolor[HTML]{EFEFEF}2}          & \multicolumn{1}{c|}{\cellcolor[HTML]{EFEFEF}3}          & \multicolumn{1}{c|}{\cellcolor[HTML]{EFEFEF}1}          & \textbf{11}                             \\ \cline{4-10} 
                                                                                                                    &                                                              & \cellcolor[HTML]{EFEFEF}                                            & \cellcolor[HTML]{EFEFEF}T                                 & \multicolumn{1}{c|}{\cellcolor[HTML]{EFEFEF}2}          & \multicolumn{1}{c|}{\cellcolor[HTML]{EFEFEF}3}          & \multicolumn{1}{c|}{\cellcolor[HTML]{EFEFEF}2}          & \multicolumn{1}{c|}{\cellcolor[HTML]{EFEFEF}3}          & \multicolumn{1}{c|}{\cellcolor[HTML]{EFEFEF}1}          & \textbf{11}                             \\ \cline{4-10} 
                                                                                                                    &                                                              & \cellcolor[HTML]{EFEFEF}                                            & \cellcolor[HTML]{EFEFEF}R                                 & \multicolumn{1}{c|}{\cellcolor[HTML]{EFEFEF}2}          & \multicolumn{1}{c|}{\cellcolor[HTML]{EFEFEF}2}          & \multicolumn{1}{c|}{\cellcolor[HTML]{EFEFEF}2}          & \multicolumn{1}{c|}{\cellcolor[HTML]{EFEFEF}3}          & \multicolumn{1}{c|}{\cellcolor[HTML]{EFEFEF}1}          & \textbf{10}                             \\ \cline{4-10} 
                                                                                                                    &                                                              & \cellcolor[HTML]{EFEFEF}                                            & \cellcolor[HTML]{EFEFEF}I                                 & \multicolumn{1}{c|}{\cellcolor[HTML]{EFEFEF}3}          & \multicolumn{1}{c|}{\cellcolor[HTML]{EFEFEF}2}          & \multicolumn{1}{c|}{\cellcolor[HTML]{EFEFEF}2}          & \multicolumn{1}{c|}{\cellcolor[HTML]{EFEFEF}3}          & \multicolumn{1}{c|}{\cellcolor[HTML]{EFEFEF}1}          & \textbf{11}                             \\ \cline{4-10} 
                                                                                                                    &                                                              & \cellcolor[HTML]{EFEFEF}                                            & \cellcolor[HTML]{EFEFEF}D                                 & \multicolumn{1}{c|}{\cellcolor[HTML]{EFEFEF}3}          & \multicolumn{1}{c|}{\cellcolor[HTML]{EFEFEF}3}          & \multicolumn{1}{c|}{\cellcolor[HTML]{EFEFEF}3}          & \multicolumn{1}{c|}{\cellcolor[HTML]{EFEFEF}3}          & \multicolumn{1}{c|}{\cellcolor[HTML]{EFEFEF}1}          & \cellcolor[HTML]{FD6864}\textbf{13}     \\ \cline{4-10} 
                                                                                                                    &                                                              & \multirow{-6}{*}{\cellcolor[HTML]{EFEFEF}Man in the middle}         & \cellcolor[HTML]{EFEFEF}E                                 & \multicolumn{1}{c|}{\cellcolor[HTML]{EFEFEF}2}          & \multicolumn{1}{c|}{\cellcolor[HTML]{EFEFEF}2}          & \multicolumn{1}{c|}{\cellcolor[HTML]{EFEFEF}2}          & \multicolumn{1}{c|}{\cellcolor[HTML]{EFEFEF}3}          & \multicolumn{1}{c|}{\cellcolor[HTML]{EFEFEF}1}          & \textbf{10}                             \\ \cline{3-10} 
                                                                                                                    &                                                              &                                                                     & I                                                         & \multicolumn{1}{c|}{3}                                  & \multicolumn{1}{c|}{1}                                  & \multicolumn{1}{c|}{3}                                  & \multicolumn{1}{c|}{3}                                  & \multicolumn{1}{c|}{1}                                  & \textbf{11}                             \\ \cline{4-10} 
                                                                                                                    & \multirow{-12}{*}{Satellite}                                 & \multirow{-2}{*}{Packet sniffing}                                   & E                                                         & \multicolumn{1}{c|}{3}                                  & \multicolumn{1}{c|}{2}                                  & \multicolumn{1}{c|}{2}                                  & \multicolumn{1}{c|}{3}                                  & \multicolumn{1}{c|}{2}                                  & \cellcolor[HTML]{FD6864}\textbf{12}     \\ \cline{2-10} 
                                                                                                                    & \cellcolor[HTML]{EFEFEF}                                     & \cellcolor[HTML]{EFEFEF}                                            & \cellcolor[HTML]{EFEFEF}S                                 & \multicolumn{1}{c|}{\cellcolor[HTML]{EFEFEF}2}          & \multicolumn{1}{c|}{\cellcolor[HTML]{EFEFEF}3}          & \multicolumn{1}{c|}{\cellcolor[HTML]{EFEFEF}2}          & \multicolumn{1}{c|}{\cellcolor[HTML]{EFEFEF}1}          & \multicolumn{1}{c|}{\cellcolor[HTML]{EFEFEF}1}          & \textbf{9}                              \\ \cline{4-10} 
                                                                                                                    & \cellcolor[HTML]{EFEFEF}                                     & \cellcolor[HTML]{EFEFEF}                                            & \cellcolor[HTML]{EFEFEF}R                                 & \multicolumn{1}{c|}{\cellcolor[HTML]{EFEFEF}2}          & \multicolumn{1}{c|}{\cellcolor[HTML]{EFEFEF}2}          & \multicolumn{1}{c|}{\cellcolor[HTML]{EFEFEF}2}          & \multicolumn{1}{c|}{\cellcolor[HTML]{EFEFEF}3}          & \multicolumn{1}{c|}{\cellcolor[HTML]{EFEFEF}2}          & \textbf{11}                             \\ \cline{4-10} 
                                                                                                                    & \cellcolor[HTML]{EFEFEF}                                     & \cellcolor[HTML]{EFEFEF}                                            & \cellcolor[HTML]{EFEFEF}I                                 & \multicolumn{1}{c|}{\cellcolor[HTML]{EFEFEF}2}          & \multicolumn{1}{c|}{\cellcolor[HTML]{EFEFEF}3}          & \multicolumn{1}{c|}{\cellcolor[HTML]{EFEFEF}2}          & \multicolumn{1}{c|}{\cellcolor[HTML]{EFEFEF}2}          & \multicolumn{1}{c|}{\cellcolor[HTML]{EFEFEF}2}          & \textbf{11}                             \\ \cline{4-10} 
                                                                                                                    & \cellcolor[HTML]{EFEFEF}                                     & \multirow{-4}{*}{\cellcolor[HTML]{EFEFEF}Spoofing}                  & \cellcolor[HTML]{EFEFEF}D                                 & \multicolumn{1}{c|}{\cellcolor[HTML]{EFEFEF}3}          & \multicolumn{1}{c|}{\cellcolor[HTML]{EFEFEF}2}          & \multicolumn{1}{c|}{\cellcolor[HTML]{EFEFEF}2}          & \multicolumn{1}{c|}{\cellcolor[HTML]{EFEFEF}3}          & \multicolumn{1}{c|}{\cellcolor[HTML]{EFEFEF}1}          & \textbf{11}                             \\ \cline{3-10} 
                                                                                                                    & \cellcolor[HTML]{EFEFEF}                                     &                                                                     & S                                                         & \multicolumn{1}{c|}{2}                                  & \multicolumn{1}{c|}{3}                                  & \multicolumn{1}{c|}{2}                                  & \multicolumn{1}{c|}{1}                                  & \multicolumn{1}{c|}{1}                                  & \textbf{9}                              \\ \cline{4-10} 
                                                                                                                    & \cellcolor[HTML]{EFEFEF}                                     &                                                                     & R                                                         & \multicolumn{1}{c|}{2}                                  & \multicolumn{1}{c|}{2}                                  & \multicolumn{1}{c|}{3}                                  & \multicolumn{1}{c|}{2}                                  & \multicolumn{1}{c|}{1}                                  & \textbf{10}                             \\ \cline{4-10} 
\multirow{-30}{*}{\begin{tabular}[c]{@{}c@{}}Inter Vehicle \\ Networks\end{tabular}}                                & \multirow{-7}{*}{\cellcolor[HTML]{EFEFEF}Vehicle to vehicle} & \multirow{-3}{*}{Fuzzing data}                                      & D                                                         & \multicolumn{1}{c|}{3}                                  & \multicolumn{1}{c|}{2}                                  & \multicolumn{1}{c|}{2}                                  & \multicolumn{1}{c|}{3}                                  & \multicolumn{1}{c|}{1}                                  & \textbf{11}                             \\ \hline
\cellcolor[HTML]{EFEFEF}                                                                                            &                                                              & \cellcolor[HTML]{EFEFEF}                                            & \cellcolor[HTML]{EFEFEF}S                                 & \multicolumn{1}{c|}{\cellcolor[HTML]{EFEFEF}2}          & \multicolumn{1}{c|}{\cellcolor[HTML]{EFEFEF}3}          & \multicolumn{1}{c|}{\cellcolor[HTML]{EFEFEF}2}          & \multicolumn{1}{c|}{\cellcolor[HTML]{EFEFEF}1}          & \multicolumn{1}{c|}{\cellcolor[HTML]{EFEFEF}1}          & \textbf{9}                              \\ \cline{4-10} 
\cellcolor[HTML]{EFEFEF}                                                                                            &                                                              & \cellcolor[HTML]{EFEFEF}                                            & \cellcolor[HTML]{EFEFEF}R                                 & \multicolumn{1}{c|}{\cellcolor[HTML]{EFEFEF}2}          & \multicolumn{1}{c|}{\cellcolor[HTML]{EFEFEF}2}          & \multicolumn{1}{c|}{\cellcolor[HTML]{EFEFEF}2}          & \multicolumn{1}{c|}{\cellcolor[HTML]{EFEFEF}2}          & \multicolumn{1}{c|}{\cellcolor[HTML]{EFEFEF}1}          & \textbf{9}                              \\ \cline{4-10} 
\multirow{-3}{*}{\cellcolor[HTML]{EFEFEF}\begin{tabular}[c]{@{}c@{}}Application \\ Layer\end{tabular}}              & \multirow{-3}{*}{Applications}                               & \multirow{-3}{*}{\cellcolor[HTML]{EFEFEF}Fuzzing data}              & \cellcolor[HTML]{EFEFEF}D                                 & \multicolumn{1}{c|}{\cellcolor[HTML]{EFEFEF}2}          & \multicolumn{1}{c|}{\cellcolor[HTML]{EFEFEF}1}          & \multicolumn{1}{c|}{\cellcolor[HTML]{EFEFEF}1}          & \multicolumn{1}{c|}{\cellcolor[HTML]{EFEFEF}3}          & \multicolumn{1}{c|}{\cellcolor[HTML]{EFEFEF}2}          & \textbf{9}                              \\ \hline
\end{tabular}
\end{table*}


\vspace{-20cm}

\section{Legal and Ethical Considerations in Autonomous Vehicle Security}\label{sec:legal}
As autonomous vehicles (AVs) become increasingly integrated into society, legal and ethical considerations around security, privacy, and data protection have become essential. Implementing robust security measures in AVs not only aims to protect users from cyber threats but also ensures compliance with existing and emerging laws and addresses the ethical implications of data collection and decision-making in AV systems. This section provides an overview of key legal and ethical challenges related to AV security.
\begin{enumerate}
    \item Data Privacy and Protection: 
    AVs collect vast amounts of data, including location information, passenger behavior, environmental data, and even biometric data in some cases. This data is crucial for AV functionality, particularly in areas like navigation, system diagnostics, and personalized services. However, handling this sensitive data raises concerns about privacy, particularly in relation to data ownership, user consent, and secure storage.
    \begin{itemize}
        \item Legal Implications: 
        Many jurisdictions now enforce stringent data protection laws, such as the European Union’s General Data Protection Regulation (GDPR) and the California Consumer Privacy Act (CCPA). These regulations require AV companies to implement strict data handling procedures, including obtaining explicit user consent, allowing users to access and control their data, and ensuring data is protected from unauthorized access. AV manufacturers must design systems to comply with these laws by default and may face significant penalties for data breaches or unauthorized data use \cite{fagnant2015preparing}.
        \item Ethical Implications: 
        Beyond legal requirements, ethical concerns about privacy arise regarding the extent to which AVs should collect personal data and how this data is used. The principle of “data minimization” suggests that AVs should only collect data that is absolutely necessary for their operation, avoiding invasive data collection that could infringe on personal privacy rights. AV developers face ethical questions about balancing the benefits of data-driven improvements against the privacy rights of individuals.

    \end{itemize}

    \item Cybersecurity and User Safety
    Ensuring the cybersecurity of AV systems is paramount, as cyberattacks on AVs can lead to severe consequences, including physical harm to passengers and pedestrians. AVs rely on interconnected systems for navigation, communication, and control, making them susceptible to a range of cyber threats, from remote hijacking to data breaches. These risks create legal obligations for AV manufacturers to safeguard users and develop adequate response protocols.
    \begin{itemize}
        \item Legal Implications: 
        Regulatory bodies in various countries are starting to introduce cybersecurity requirements specifically for AVs. For example, the United Nations Economic Commission for Europe (UNECE) introduced WP.29, a regulatory framework that mandates cybersecurity measures in vehicles to prevent and mitigate cyberattacks. These regulations require manufacturers to adopt "cybersecurity by design" approaches, ensuring that security measures are integrated into the AV development process and continuously updated to protect against evolving threats \cite{regulation2021uniform}.
        \item Ethical Implications: 
        Ethically, AV developers must consider the duty of care owed to passengers and others impacted by AV operations. Security vulnerabilities that compromise passenger safety raise ethical questions about the responsibility of AV developers to ensure robust protection against attacks. There is an ethical obligation to prioritize user safety and security, even if these measures increase development costs or reduce product launch timelines.
    \end{itemize}

    \item Transparency and Accountability in Decision-Making
    AVs make complex, autonomous decisions, particularly in dynamic driving environments, which raises ethical concerns about the transparency and accountability of these decisions. For example, if an AV is involved in an accident, it is essential to determine responsibility and understand the decision-making process that led to the incident.
    \begin{itemize}
        \item Legal Implications: 
        Current laws struggle to address liability in situations where AVs make autonomous decisions, especially in “trolley problem” scenarios where the AV must choose between two potentially harmful actions. Some jurisdictions are working to establish liability frameworks, specifying whether responsibility should lie with the manufacturer, software developer, or vehicle owner. These liability questions are crucial for defining who is accountable when security breaches or system failures lead to accidents \cite{hevelke2015responsibility}.
        \item Ethical Implications: 
        Ethically, AV systems should operate transparently, with clear, understandable explanations for their decisions, especially in critical scenarios. The use of AI models for decision-making raises questions about the fairness and bias in these systems, which may affect people differently based on demographic factors. Developers face the ethical responsibility of ensuring that AV algorithms are fair, unbiased, and interpretable.
    \end{itemize}

    \item Legal Compliance with Safety Standards
    As AVs operate in public spaces and interact with other vehicles, pedestrians, and infrastructure, they must meet specific safety standards to ensure they do not pose undue risks to others. These standards often include compliance with technical and operational safety requirements, such as those enforced by the U.S. National Highway Traffic Safety Administration (NHTSA) and similar agencies worldwide.
    \begin{itemize}
        \item Legal Implications: 
        Compliance with safety standards is mandatory for AV deployment, and manufacturers can face legal repercussions if their vehicles fail to meet these standards. Standards for AV safety often cover both functional safety (preventing system failures) and safety of the intended functionality (ensuring the AV performs safely under all circumstances). These standards are essential to prevent accidents, ensure reliable performance, and maintain public trust in AV technologies \cite{owens2020notice}.
        \item Ethical Implications: 
        Ethically, AV developers are responsible for ensuring that their products prioritize safety at every level of design and functionality. Meeting safety standards is not just a legal requirement but also an ethical commitment to protect the lives of users and the general public. AV developers must weigh the potential risks of their designs and actively work to mitigate them, even if this involves delaying deployment or increasing development costs.
    \end{itemize}
\end{enumerate}
The legal and ethical considerations surrounding AV security highlight the complexity of developing autonomous systems that are safe, secure, and respectful of user privacy. Addressing these issues requires a collaborative approach between AV developers, policymakers, and ethicists to establish frameworks that ensure robust security measures, accountability, and respect for personal privacy. As the field of AV technology continues to advance, so too must the legal and ethical standards guiding its responsible development and deployment.

\section{Future Direction and Discussion} \label{sec:future}
As autonomous vehicle (AV) technologies continue to advance, ensuring robust cybersecurity remains a critical challenge. Emerging research areas such as blockchain for Vehicle-to-Everything (V2X) security, AI-driven threat detection, and secure Over-The-Air (OTA) update mechanisms offer promising solutions to address the dynamic threat landscape. This section explores these directions and provides actionable recommendations for AV developers and policymakers to enhance AV security frameworks.

\subsection{Blockchain for V2X Security}
Blockchain technology presents a viable solution for securing V2X communication by offering a decentralized and tamper-resistant ledger. This approach enhances the authenticity and integrity of messages exchanged between vehicles (V2V), infrastructure (V2I), and networks (V2N), mitigating risks associated with message spoofing and data manipulation \cite{kukkala2022roadmap, ahmad2024machine}. For instance, blockchain-based secure message authentication protocols can verify the origin and integrity of transmitted data, preventing malicious actors from injecting false information into the network\cite{shangguan2022dynamic}. Recent studies have shown that hybrid blockchain models, which combine public and private blockchains, can achieve a balance between security, scalability, and latency\cite{saeed2023review}.\\
Blockchain also offers decentralized trust management, reducing reliance on central authorities and eliminating single points of failure\cite{potteiger2016software}. Immutable communication logs maintained by blockchain provide an auditable trail, facilitating forensic investigations in the event of security breaches\cite{kwon2020cyber}. To achieve widespread adoption, standardized blockchain protocols for V2X communication need to be developed to ensure interoperability across different AV manufacturers and infrastructure providers\cite{nie_blackhat}. Additionally, pilot implementations in controlled environments are necessary to evaluate the performance, scalability, and security of blockchain-based V2X systems\cite{eiza2017driving}.

\subsection{AI-Driven Threat Detection and Mitigation}
Artificial Intelligence (AI) and Machine Learning (ML) play an essential role in enhancing AV security by enabling real-time threat detection and mitigation \cite{lim2024impact}. AI-driven models can analyze large datasets from sensors, communication networks, and onboard systems to identify anomalies indicative of cyberattacks\cite{wu2023susceptibility}. For example, AI-based Intrusion Detection Systems (IDS) can detect deviations in network traffic, helping to identify threats such as sensor spoofing, CAN bus injection, and Denial-of-Service (DoS) attacks \cite{khan2020cyber, hamad2020savta}. Furthermore, AI models can be trained to withstand adversarial attacks, enhancing the resilience of AV perception systems by cross-validating inputs from multiple sensors, including LiDAR, radar, and cameras\cite{girdhar2023cybersecurity}.\\
Recent research highlights the importance of federated learning for distributed AI training across AV fleets, allowing vehicles to collaboratively improve threat detection models without sharing raw data\cite{khayyam2020artificial}. Predictive security models that use historical data and emerging threat trends can help anticipate vulnerabilities and proactively implement mitigation measures\cite{ma2020artificial}. Additionally, the integration of reinforcement learning techniques can enhance adaptive security mechanisms, enabling AVs to respond dynamically to evolving threats\cite{petit2015cyberattacks}.\\
To effectively leverage AI for AV security, developers should integrate AI-driven IDS into vehicle networks, develop resilient perception algorithms that combine multi-sensor data, and invest in federated learning to enhance collaborative threat detection\cite{xia2021survey}.

\subsection{Secure OTA Update Mechanisms}
Secure Over-The-Air (OTA) update mechanisms are essential for maintaining the integrity and security of AV software and firmware\cite{szymonik2024cybersecurity}. OTA updates enable AV manufacturers to deploy patches, fix vulnerabilities, and introduce new functionalities. However, insecure OTA processes can be exploited to inject malicious code or disrupt vehicle operations\cite{tencent2020}. To mitigate these risks, end-to-end encryption should be employed to protect updates from interception and tampering \cite{tabani2021performance}. Recent advancements in blockchain technology also offer a promising approach for securing OTA updates by maintaining immutable records of update transactions\cite{sharma2017distblocknet}.
\\
Code-signing and integrity verification mechanisms should be used to authenticate update packages before installation, ensuring that only verified updates are applied\cite{schmittner2016using}. Additionally, secure boot processes should be implemented to verify firmware integrity during startup, preventing unauthorized or corrupted software from running\cite{cena2023composite}. Rollback mechanisms can be employed to revert to a previous stable state if an update is compromised or fails\cite{hussain2019autonomous}.
\\
Regular audits and vulnerability assessments of OTA systems are necessary to identify and address potential security weaknesses\cite{halder2020secure}. Policymakers should mandate secure OTA update practices through regulations such as ISO/SAE 21434 and UNECE WP.29, which outline cybersecurity requirements for automotive software updates\cite{UNECE2021}.

\subsection{Actionable Recommendations for AV Developers and Policymaker}
\begin{itemize}
    \item \textbf{Develop Regulatory Frameworks for AV Security:} Policymakers should establish guidelines that mandate secure V2X communication, AI-based threat detection, and secure OTA update mechanisms. Regulations such as ISO/SAE 21434 and UNECE WP.29 should be enforced to ensure comprehensive cybersecurity practices\cite{hamad2020savta}.
    \item \textbf{Incentivize Cybersecurity Research:} Governments should provide funding and grants to support research in blockchain, AI, and secure software development for AVs\cite{petit2015cyberattacks}. Collaborative research initiatives between academia, industry, and government agencies can accelerate innovation and address emerging threats\cite{folan2023cybersecurity}.
    \item \textbf{Promote Industry Collaboration:} Encourage partnerships between AV manufacturers, cybersecurity experts, and academic institutions to develop and share best practices for AV security\cite{linkov2019human}. Establishing open forums and working groups can facilitate the exchange of knowledge and strategies \cite{miller2015remote}.
    \item \textbf{Public Awareness and Education:} Launch educational campaigns to inform consumers about AV security features and the importance of regular software updates. Increased awareness can drive demand for secure AV technologies and encourage manufacturers to prioritize cybersecurity.
\end{itemize}
By focusing on these research areas and implementing these recommendations, the security and resilience of autonomous vehicles can be significantly enhanced. These advancements will help mitigate cybersecurity risks, protect user safety, and ensure public trust in the deployment of AV technologies.

\section{Conclusions} \label{sec:conc}
Autonomous vehicles (AVs) represent a transformative shift in modern transportation, promising increased safety, efficiency, and convenience. However, the growing complexity and interconnectedness of AV systems introduce a broad range of cybersecurity vulnerabilities that must be addressed to ensure safe and reliable deployment. This paper has provided an in-depth exploration of the threats and security challenges faced by AVs, focusing on threat modeling frameworks such as STRIDE, DREAD, and MITRE ATT\&CK to systematically identify and mitigate potential risks.
\\
Key findings highlight that AVs are susceptible to various attack vectors, including wireless communication exploits, sensor spoofing, and firmware vulnerabilities. Case studies of real-world incidents, such as the Jeep Cherokee and Tesla exploits, underscore the critical need for robust security measures. Threat modeling frameworks like STRIDE and DREAD offer structured approaches to categorize threats and prioritize mitigation efforts, while the MITRE ATT\&CK framework provides detailed insights into adversarial tactics and techniques.
\\
Emerging technologies, including blockchain for V2X communication security, AI-driven threat detection, and secure Over-The-Air (OTA) update mechanisms, show significant promise in enhancing AV cybersecurity. These innovations, combined with rigorous threat modeling and multi-layered defense strategies, are essential for protecting AV systems against evolving cyber threats.
\\
Legal and ethical considerations, such as data privacy, user safety, and transparency in decision-making, play a crucial role in guiding the development and deployment of AVs. Compliance with regulations like GDPR, ISO/SAE 21434, and UNECE WP.29 is necessary to address these concerns and ensure public trust in AV technology.
\\
Looking ahead, collaboration between AV developers, cybersecurity experts, policymakers, and researchers is vital for advancing secure AV technologies. Continued research into adaptive security solutions, adversarial AI defenses, and secure communication protocols will be critical for addressing the dynamic threat landscape. By adopting proactive threat modeling and integrating cutting-edge security measures, the AV industry can pave the way for safer and more resilient autonomous transportation systems.
\\
In conclusion, ensuring the cybersecurity of autonomous vehicles requires a comprehensive, multi-faceted approach that combines technical, legal, and ethical considerations. As AV technologies continue to evolve, our strategies for protecting them must also evolve, ensuring that the benefits of autonomous transportation are realized safely and securely.

\bibliographystyle{IEEEtran}
\bibliography{ref}

\end{document}